\documentclass[prd,aps,nofootinbib,preprintnumbers]{revtex4}
\usepackage{amsmath,amssymb,epsf}
\usepackage{graphicx, pstricks}
\def\be{\begin{equation}}
\def\ee{\end{equation}}
\def\ba{\begin{equation}}
\def\ba{\begin{eqnarray}}
\def\ea{\end{eqnarray}}

\def\bq{\begin{quote}}
\def\eq{\end{quote}}

\def\hlf{\frac{1}{2}}

\def\lmk{\left(}
\def\rmk{\right)}

\def\hlf{\frac{1}{2}}
\begin{document}

\title{Preheating with Trilinear Interactions: Tachyonic Resonance}
\author{J.F. Dufaux$^{1}$, G.N. Felder$^{2}$, L. Kofman$^{1}$,
M. Peloso$^{3}$, D. Podolsky$^{1}$\footnote{On leave from Landau
Institute for Theoretical Physics, 117940, Moscow, Russia}}
\affiliation{${}^1$CITA, University of Toronto, 60 St. George st.,
Toronto, ON M5S 3H8, Canada}
\affiliation{${}^2$Department of Physics, Clark Science Center, Smith College 
Northampton, MA 01063, USA}
\affiliation{${}^3$School of Physics and Astronomy, University of 
Minnesota, Minneapolis, MN 55455, USA}
\date{\today}

\begin{abstract}
We investigate the effects of bosonic trilinear interactions in  preheating after
chaotic inflation. A trilinear interaction term allows for the complete decay of
the massive inflaton particles, which is necessary for the transition to radiation
domination. We found that typically the trilinear term is subdominant during early
stages of preheating, but it actually amplifies parametric resonance driven by the
four-legs interaction. In cases where the trilinear term does dominate
during preheating, the process occurs through periodic tachyonic amplifications with 
resonance effects, which is so effective that preheating completes
within a few inflaton 
oscillations. We develop an analytic theory of this process, which we call tachyonic resonance. 
We also study numerically the influence of trilinear interactions on the dynamics after preheating. 
The trilinear term eventually comes to dominate after preheating, leading to faster rescattering and
thermalization than could occur without it. Finally, we investigate the role of non-renormalizable 
interaction terms during preheating. We find that if they are present they
generally dominate (while still in a controllable regime) in
chaotic inflation models. Preheating due to these terms proceeds through a modified
form of tachyonic resonance.
\end{abstract}

\preprint{UMN-TH-2431/06}
\maketitle

\section{Introduction }
\label{introduction}

According to the inflationary scenario, the universe at early times
expands quasi-exponentially in a vacuum-like state without entropy or
particles. During this stage of inflation, all energy is contained in
a classical slowly moving inflaton field.  Eventually  the
inflaton field decays and transfers all of its energy to relativistic
particles, which starts the thermal history of the hot Friedmann universe.
Particle creation in preheating, described by quantum field theory, is
a spectacular process where all the particles of the universe are created
from the classical inflaton. 
 In chaotic inflationary models,
soon after the end of inflation the almost homogeneous inflaton field
 coherently oscillates with a very large amplitude of the
order of the Planck mass around the minimum of its potential.  Due to its 
interactions with other fields, the
inflaton decays and transfers all of its energy to relativistic
particles.  If the creation of particles is sufficiently slow (for
instance, if the inflaton is coupled only gravitationally to the
matter fields) the decay products simultaneously interact with each
other and come to a state of thermal equilibrium at the reheating
temperature $T_r$. This gradual reheating can be treated with the
perturbative theory of particle creation and thermalization~\cite{fermions}.  
However, for a wide range of couplings
the particle production from a
coherently oscillating inflaton occurs in
the non-perturbative regime of parametric
excitation~\cite{kls1, KLS97, traschen, shtanov}.  
This picture, with variation in its
details, is extended to other inflationary models. For instance, in
hybrid inflation (including $D$-term inflation) the inflaton decay
proceeds via a tachyonic instability
of the inhomogeneous modes which accompany the symmetry breaking~\cite{tachyonic}.  
One consistent feature of preheating -- non-perturbative copious
particle production immediately after inflation -- is that the process
occurs far away from thermal equilibrium.
The transition from this stage  to thermal equilibrium  occurs in 
a few distinct stages, each
much longer than the previous one. First there is the rapid
preheating phase, followed by the onset of turbulent 
interactions between the different modes. Our understanding of this stage comes 
from lattice numerical simulations~\cite{lattice} as well as from different 
theoretical techniques~\cite{berges}. For a wide range
of models, the dynamics of scalar field turbulence is largely independent of the
details of inflation and preheating~\cite{fk}. Finally, there is
thermalization, ending with  equilibrium. In general, the equation of
state of the universe is that of matter when it is dominated by the
coherent oscillations of the inflaton field, but changes when the
inflaton decays into radiation-dominated plasma~\cite{podol}.

Most studies of preheating have focused on the models with
  $\phi^2\chi^2$
four-legs interactions of the inflaton $\phi$ with another scalar field $\chi$.
 A common feature of 
preheating is the production of a large number of inflaton 
quanta with non-zero momentum from rescattering, alongside with
inflatons at rest.
 The momenta of these relic massive inflaton particles eventually
would redshift out.
However, the
decay of inflaton particles through four-legs $\phi\phi \rightarrow \chi\chi$
processes in an expanding universe is never complete.
 Thus inflaton particles  later on will have a matter equation of state
 and  come to dominate the energy density, which is not an acceptable scenario.
 Therefore, to avoid this, we must include in the theory of reheating
 interactions of the type $\phi \chi^n \,$, that allow the inflaton to decay completely, thus 
resulting in a radiation dominated stage. Trilinear interactions are the most immediate and natural
interactions of this sort.

Trilinear interactions occur commonly in many theoretical models.
Yukawa couplings, for example, lead to trilinear vertices with fermions.
Three-legs decay via fermions was in fact the first channel of inflaton decay
considered in early papers on inflation~\cite{fermions} within perturbation theory.
It was later realized that even in the case of interaction with fermions the inflaton decay typically 
 occurs via non-perturbative parametric excitations~\cite{excit}.
However, three-legs decay via intractions with bosons is expected to
 be a dominant channel (due to Pauli blocking for fermions).
Gauge interactions lead to trilinear vertices with vector fields. 
Even if we restrict ourselves to scalar field interactions,  trilinear
interactions naturally appear in many contexts.

Consider for instance a chaotic inflation model with the effective potential 
$V =\pm \frac{1}{2}m^2 \phi^2 +\frac{1}{4}  \lambda \phi^4 + \frac{g^2}{2} \, \phi^2 \, \chi^2 $.
The negative sign corresponds to spontaneous symmetry breaking 
${\phi} \to \phi +\sigma \,$, which results in a classical scalar
field VEV $\sigma=\frac{m}{\sqrt{\lambda}}$. The interaction term $g^2 \phi^2 \chi^2$ 
then gives rise to a trilinear vertex $\sigma g^2\, {\phi} \, \chi^2$ 
along with the four-legs interaction. Spontaneous symmetry breaking naturally 
emerges in the new inflationary scenario. However, this results in a mass for 
the $\chi$ particles which is comparable to the inflaton mass. This complicates the 
bosonic decay of the inflaton after new inflation, see also~\cite{Desroche:2005yt}.
One encounters the same problem for spontaneous symmetry breaking in chaotic inflation models. 
However, a very different picture appears for chaotic inflation in supersymmetric theories. 

Trilinear interactions occur also naturally in
supersymmetric theories. Typical superpotentials
are constructed from trilinear combinations of superfields,  
$W=\sum_{ijk} \lambda_{ijk} \Phi_i \Phi_j \Phi_k$. 
The cross terms in the potential
$V = \sum_i \vert { W_i} \vert^2$ then give rise to a three-legs interaction, in addition 
to a four-legs vertex.
Besides many other advantages, 
supersymmetry is useful to protect the flatness of the inflaton
potential from large radiative corrections.
Therefore we will pay special attention to
trilinear interactions in SUSY theories.
In SUSY the strengths of the couplings of three-legs and four-legs
interactions are tightly connected.

The scalar field potentials that we consider in this paper, and their stability constraints,  
are discussed in Section~\ref{3pot}. 

In general, we expect the inflaton to decay simultaneously 
via all potential interactions: three-legs, four-legs etc.
In many cases, notably in the  supersymmetric case,
 the leading channel of inflaton decay during  preheating is 
bosonic four-legs parametric resonance, as it was assumed in the earlier papers
on preheating \cite{KLS97}. In this case, parametric resonance is actually amplified 
by the trilinear interaction, as we show in Section~\ref{sec:comparison}.

To reach this important conclusion, we shall understand how preheating works 
with a three-legs interaction. For this
 we shall consider an idealized situation when
other interactions are switched off.

To get  an insight into how preheating  occurs through a trilinear vertex
$\frac{1}{2}\sigma \phi \chi^2$, consider 
 the effective frequency  of $\chi$ particles caused by the 
 inflaton field $\phi(t) =\Phi\,\sin mt$ coherently oscillating around the origin
\begin{equation}\label{freq}
\omega_{\chi}^2 ( t)=\frac{k^2}{a^2}+ m_\chi^2 + \sigma \, \phi \left( t \right) \ .
\end{equation}
The bare mass of the $\chi$ will typically satisfy
$m_{\chi}^2 << \sigma\,\Phi$. The trilinear coupling then leads to a
tachyonic mass for $\chi$ whenever $\phi < 0$, 
 which  happens during half of each inflaton oscillation. Correspondingly,
the  modes with $k^2 < \sigma\, |\phi|\, a^2$  will be exponentially
amplified during a portion of each half-period of the inflaton oscillations.
Therefore  preheating arising from a
trilinear vertex shares features of both parametric
resonance and tachyonic preheating, so we call this process {\it tachyonic
resonance}. Such tachyonic resonance  leads to very efficient production of 
$\chi\,$ particles, so that preheating in this case concludes within the first
few oscillations of $\phi \,$.

Aspects of parametric resonance in the presence of a trilinear interaction 
appeared already in~\cite{shtanov}.  
Tachyonic resonance effects may be recognised with our present understanding in 
a numerical plot of~\cite{Garcia-Bellido:1999sv} in the context of preheating after 
hybrid inflation. More generally, combined tachyonic and resonant effects appeared 
in the context of dynamical symmetry breaking \cite{tachyonic},
and after new inflation~\cite{Desroche:2005yt}.
The tachyonic effect from a trilinear interaction
was  studied in the context of the cosmological moduli problem \cite{shbr}.
In this paper, we systematically construct the theory of tachyonic resonance from
trilinear interactions in preheating after chaotic inflation.

The origin of tachyonic resonance is similar to that of parametric resonance.
Without expansion of the universe $(a(t)=1)$,
they both can be understood  using the stability/instability chart of the Mathieu
equation. Introduce dimensionless parameters entering the frequency
(\ref{freq}), $A=\frac{k^2}{m^2}$, $q=\frac{\sigma \Phi_0}{m^2}$
(as customary  for the theory of preheating). The 
 $(A,q)$ half-plane ($A>0$) is  divided 
into stability and instability stripes (see, e.g. \cite{abram}).
The line $A=2q$ divides the stability/instability chart into two regions.
 The case of parametric resonance 
for  the four-legs interaction $g^2 \phi^2 \chi^2$ 
can be mapped onto the theory of stability/instability chart above the line $A=2q$
studied in \cite{KLS97}. The region below that line
corresponds to what we call tachyonic resonance.
On the formal side,  we will extend the theory \cite{KLS97} of parametric resonance
``below the line $A=2q$'', which requires some new technical
elements.
The analytic theory of trilinear preheating will be
constructed  in Section~\ref{tachres}.

In this paper, we study the effects of the trilinear $\phi\;\chi^2$
interaction both during preheating and during
the rescattering/early thermalization stage.
Preheating produces a spectrum of particles predominantly in the
infrared (low momentum) region. While $2 \leftrightarrow 2$ interactions can lead to kinetic
equilibrium, thermal equilibrium requires an increase of the average
energy per particle, which in turn requires particle fusion. The
beginning of particle fusion is already visible during rescattering,
as the simulations of~\cite{fk} indicate. However, this process is
expected to complete on time scales which are much longer than the
ones we can run in a lattice simulation. Thermalization after
preheating has been studied in detail only in limited intervals of time
and momentum \cite{mt,fk,podol}.
An analytic treatment of the dynamics of thermalization is possible 
 only in the simplest cases, for
instance,  in the pure $\lambda \, \phi^4$ model~\cite{mt},
where the expansion of the universe can be scaled out and no mass scale
appears. In an expanding universe in the more general case
of a massive inflaton it is much harder to do analytically \cite{fk,podol}.
 If only quartic interactions are relevant,
particle fusion proceeds most effectively by combining two vertices in
a $4 \rightarrow 2$ process. If trilinear vertices are also present
with comparable coupling strength, $3 \rightarrow 2$ processes
become possible. Hence,
trilinear vertices can be expected to play an important role in setting
the thermalization time scale. Yukawa couplings can also be important for thermalisation, 
see e.g.~\cite{Aarts:1998td}~\footnote{Trilinear interactions may also play an important role 
for electroweak baryogenesis and leptogenesis, see~\cite{Garcia-Bellido:2003wd} and references 
therein}. In Section~\ref{numsimul}, we
present the results of our numerical simulations and we discuss
the effect of the trilinear interaction on the dynamics after
preheating. We focus in particular on the
evolution of the equation of state.

As we already mentioned, during preheating four-legs interactions will
often dominate
over three-legs interactions. In many cases we also may expect the presence of higher-order
interactions like $\frac{1}{M} \chi^2 \phi^3$, $\frac{1}{M^2}\chi^2 \phi^4$, etc.
 These non-renormalizable terms
arise for example in supergravity theories. 
The  cut-off scale $M$ of the effective theory is expected to be close to the Planck scale.
 Planck-suppressed interactions are often supposed to lead
only to perturbative contributions to reheating.
The situation, however,  may be very
different for chaotic inflation because of the large inflaton amplitude 
 right after inflation, $\Phi \sim 0.1\;M_P$.
 (Clearly, one has also to make sure that non-renormalizable
terms do not affect the inflaton potential during inflation).
In fact, surprisingly, non-renormalizable terms tend to dominate over renormalizable terms 
during preheating, as long as $g^2 < \frac{\Phi}{M}$, where $g^2$ is the coupling of the 
four-legs interaction.

For $M$ close to the Planck mass, $\Phi/M < 1$ during preheating, and 
the effective mass squared of $\chi$ is then actually dominated by the $\phi^3 \, \chi^2 / M$
interaction. This vertex leads to tachyonic growth of the quanta of
$\chi$, similar to what occurs with a three legs interaction. Note that the
efficiency of this process is governed by the dimensionless parameter
$q \sim \Phi^3/(m^2\,M)$, which is very large at the end of
inflation. This also opens up the possibility of very efficient
preheating even if the inflaton does not couple to the matter sector
through renormalizable interactions\footnote{If a Bose condensate associated with 
a flat direction forms during inflation, non-renormalizable interactions may also play 
an important role during preheating, as noticed in~\cite{Allahverdi:2004ge}.}. 
We consider tachyonic resonance due to 
non-renormalizable interactions in Section~\ref{nonrenorm}.

 We briefly summarize our results in 
section \ref{conclusions}.

\section{Properties of Potentials With Trilinear Interactions}
\label{3pot}

In this section, we describe the models that we will consider in the following, and which are characterized by a trilinear 
interaction $\phi \chi^2$ between the massive inflaton $\phi$ and a light matter scalar field $\chi \,$.
In the presence of a $\phi\chi^2$ vertex (independently of the presence of a $\phi^2\chi^2$ interaction), we have to 
include a 
$\chi^4$ self interaction as well for the potential to be bounded from below, and therefore stable. This means that
there are several effects of the fields dynamics working simultaneously.

We consider two  examples of potentials with three-legs interactions,
which stress different possibilities of preheating.

\begin{figure}[hbt]
\begin{center}
\includegraphics[height=7cm]{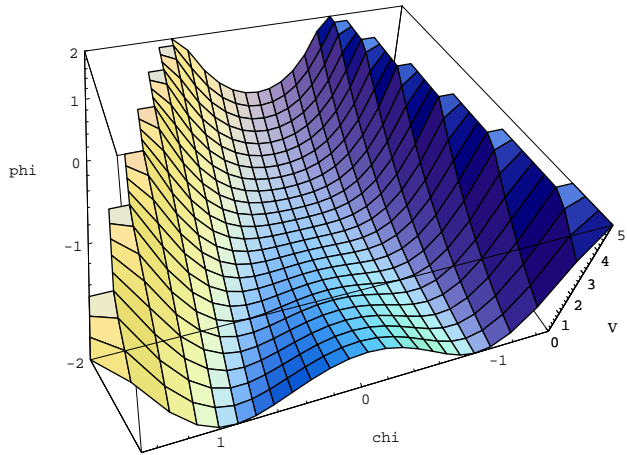}
\caption{The potential (\ref{v3}) for $\lambda = \sigma^2/2m^2$ with an exact flat direction. 
For $\lambda > \sigma^2/2m^2$ the flat direction is lifted and the single minimum occurs 
at $\phi = \chi = 0$. Inflation occurs as $\phi$
rolls down the single valley at the top (positive $\phi$) and
preheating occurs as it oscillates between the single-valley and the
double-valley.}
\label{vplot}
\end{center}
\end{figure}

1) A simple  potential which encodes the trilinear interaction between the massive inflaton $\phi$ 
and another scalar field $\chi$ is supplemented with a quartic self-interaction of $\chi$ 
but without a  four-legs  interaction $\phi^2 \chi^2$
\begin{equation}
\label{v3}
V = \frac{m^2}{2}\,\phi^2 + \frac{\sigma}{2}\,\phi\,\chi^2 + \frac{\lambda}{4}\,\chi^4
\end{equation}
where $\sigma$ has the dimension of mass. Here we assume that the bare mass of $\chi$ 
is negligible with respect to the inflaton mass. 
It is instructive  to rewrite (\ref{v3}) in the form
\begin{equation}
V = \frac{1}{2}\,\left(m\,\phi + \frac{\sigma}{2m}\,\chi^2\right)^2 + 
\frac{1}{4}\,\left(\lambda - \frac{\sigma^2}{2m^2}\right)\,\chi^4 \ .
\end{equation}
Values $\lambda < \sigma^2/2m^2$ are not allowed, since in this case the potential is not bounded 
from below. The limiting case $\lambda = \sigma^2/2m^2$ is characterized by the exact flat direction 
$\phi = -\sigma\chi^2/2m^2$ where $V = 0$. The shape of the potential in this case is shown in 
Fig.\ref{vplot}. Finally, for $\lambda > \sigma^2/2m^2$ the flat direction is lifted and the potential 
admits a single minimum at $\phi = \chi = 0$ where $V = 0$. For numerical estimates, we will often take 
$\lambda = \sigma^2/m^2$ in the following. 

The inflationary stage is driven by the $\phi$-field in (\ref{v3}) if the
effective mass of the $\chi$ field is large, i.e. if 
$\sigma\,\phi(t) > H^2> m^2$ during inflation, and the $\chi$ field then stays 
at the minimum of its effective potential, at $\chi = 0$.
After inflation  $\phi$ begins to oscillate
 around the minimum. We shall  consider creation of 
 particles of the test quantum field $\chi$
 due to its trilinear interaction with the
 background oscillations of the classical field $\phi$.

When the inflaton passes the minimum of the
potential, $\sigma\,\phi(t)$ becomes negative and $\chi$
acquires a negative mass squared around  the
local maximum in the $\chi$-direction at $\chi = 0$ (see Figure~\ref{vplot}).
As a result
inhomogeneous quantum fluctuations of $\chi$ with momenta $k^2 <
|m^2_{\chi, \mathrm{eff}}(t)| = |\sigma\,\phi(t)|$ grow exponentially
with time, in a way similar to
 the tachyonic instability in hybrid
inflation discussed in \cite{tachyonic}.
The difference here, however, is that tachyonic instability occurs
only during the periods when the background field $\phi$ is negative
and ceases during the parts of the $\phi$ oscillations when $\phi>0$.

2) We will also consider the potential
\be
\label{v4}
V = \frac{m^2}{2}\,\phi^2 + \frac{\sigma}{2}\,\phi\,\chi^2 + \frac{g^2}{2}\,\phi^2\,\chi^2  
+ \frac{\lambda}{4}\,\chi^4
\ee
with the additional four legs interaction. The potential is bounded from below for $\lambda > 0$. 
For $0 < \lambda < \sigma^2/2m^2$, the 
potential admits two minima at non vanishing $\phi$ and $\chi$ where $V < 0$.
We mostly  focus on the case 
 $\lambda \ge \sigma^2/2m^2$, where there is a single minimum at $\phi = \chi = 0$ with $V = 0$. Close 
to this minimum the shape of the potential is similar to (\ref{v3}).

 The potential (\ref{v4}) may arise for instance from the superpotential
\be
\label{w4}
W = \frac{m}{2\sqrt{2}}\,\phi^2 + \frac{g}{2\sqrt{2}}\,\phi\,\chi^2
\ee
which gives $\lambda = g^2/2$ and $\sigma = g m$ (for the real part of $\chi$). Preheating 
in the theory (\ref{v4}) 
occurs through a combination of effects due to the $\phi^2\,\chi^2$ interaction and 
 the $\phi\,\chi^2$ interaction.
In the  Section \ref{sec:comparison} we discuss the range of
parameters for which the different interactions dominate.

\section{Analytical Theory of Tachyonic Resonance}
\label{tachres}

In this section, we study the production of quanta of a scalar field
$\chi$ whose effective mass squared periodically becomes negative in
time due to its interaction with the classical inflaton field
oscillating around its minimum. For definiteness, we consider the
potential (\ref{v3}). The backreaction of the $\chi^4$
self-interaction is negligible during the early stage of preheating,
so that only the $\phi \chi^2$ term is important.

The quantum field $\hat \chi$  is decomposed
into creation and annihilation operators with the eigenmodes $\chi_k(t) \, e^{i {\bold k} \vec x}$,
where $ {\bold k}$ is the co-moving momentum. 
After the usual rescaling  $\chi_k \rightarrow a^{3/2}\,\chi_k$, where $a(t)$ is the scale factor of the 
(spatially flat) FRW universe, the temporal
part of the $\chi$-modes obeys the equation \be
\label{modes}
\ddot{\chi}_k + \omega_k^2\;\chi_k = 0 \ ,
\ee
where
\begin{equation}
\label{omexp}
\omega_k^2 = \frac{k^2}{a^2} + \sigma\,\phi(t) + \Delta \ ,
\end{equation}
$\Delta = -3\dot{a}^2/4a^2 -3\ddot{a}/2a$, and a dot denotes a
derivative with respect to the proper time $t$. 
The choice of the positive frequency asymptotic solution in the past fixes the initial conditions
while the condition $\chi_k \dot{\chi}_k^* - \chi_k^* \dot{\chi}_k = i$
fixes the normalization for the solution of equation (\ref{modes}).
In this section we consider solutions of this equation for harmonically oscillating $\phi(t)$.

Let us  first neglect the expansion of the universe, $a(t) = 1$. The background solution for the inflaton 
is then given by $\phi(t) = \Phi\,\sin(mt)$ with a constant amplitude $\Phi$. In this case, Eq.(\ref{modes}) 
reduces to the canonical Mathieu equation  
\be
\chi_k '' + (A_k - 2q \cos 2z) \chi_k = 0
\label{mathieu}
\ee
where $mt = 2 z - \frac{\pi}{2}$, $A_k = \frac{4k^2}{m^2}$ and $q = \frac{2\sigma \Phi}{m^2}$, and 
a prime denotes the derivative with respect to $z$. Right after inflation we have $\Phi_0 \sim 0.1\,M_P$, 
while $m \sim 10^{13}$ GeV in order to match with the observed CMB anisotropies. The $q$-parameter is 
then  very large, for instance $q_0 \sim \sqrt{\lambda}\,10^5 >> 1$ for $\lambda \sim \sigma^2/m^2$. 

The stability/intsability chart $(A, q)$ of the Mathieu equation is divided into bands. 
We are interested in unstable solutions, which correspond to the amplification of the vacuum fluctuations $\chi_k$,
in the regime of large $q$. Let us inspect the effective frequency $\omega_k^2$.
Modes with momenta $A_k \ge 2 q$ have positive  $\omega_k^2$.
We can view Eq.(\ref{modes}) as the Schrodinger equation with the periodic potential,
where for $A_k \ge 2 q$ the waves  propagate above  the potential
barrier and are periodically scattered by its peaks.
They are amplified at the instances when $\omega_k^2$ is minimal, i.e. 
 in the vicinity of $z = l \pi$ which corresponds to $\phi = - \Phi$ in our
 case, and remain adiabatic outside of those instances. The whole process can be considered
as a series of scatterings in a parabolic potential, which approximates  $\omega_k^2$
around its zeros. The net effect corresponds to broad parametric resonance, as described in \cite{KLS97}.

However, for  $0 < A_k < 2q$
the frequency squared of the
 modes with $k^2 < \sigma\,\Phi$ becomes  negative during some 
intervals of time within each period of the background oscillations.
These are the modes we consider in the rest of this section because they give the dominant 
contribution to the number of produced particles. 
For those, we still can use the method of successive scatterings,
but each individual scattering cannot  be approximated as scattering
from a parabolic potential.
 This is where we have to modify the theory.

Suppose the inflaton oscillations  begin at some initial time $t=t_0$ where $\omega_k^2(t_0) > 0$ (take
$\phi(t_0) = \Phi$). Consider the period during the $(j+1)^{\mathrm{th}}$ oscillation of the inflaton, 
from $t = t_j$ to $t = t_{j+1}$ where $t_j = t_0 + 2\pi\,j/m$. The frequency squared of the modes with 
$k^2 < \sigma\,\Phi$ is negative during a $k$-dependent interval
\be
\Omega_k^2(t) = - \omega_k^2(t) > 0  \;\;\; \mbox{ for } \;\;\; t_{kj}^- < t < t_{kj}^+
\ee
between the ``turning points'' $t_{kj}^-$ and $t_{kj}^+$ where $\omega_k^2 = 0$.
Now we can view Eq.(\ref{modes}) with $0 < A_k < 2q$
 as the Schrodinger equation with a periodic potential,
where the wave periodically propagates above and below the potential barrier.

Except in the vicinity of the 
turning points, we have $|\dot{\omega}_k| << |\omega_k|^2$ and $|\ddot{\omega}_k| << |\omega_k|^3$ (including 
imaginary values for $\omega_k$) for $q >> 1$, so we may use the WKB approximation to solve for (\ref{modes}). 
For $t < t_{kj}^-$ (above the barrier), we have a  superposition of positive- and negative-frequency waves
\be
\label{chiom}
\chi_k(t) \simeq \chi_k^j(t) = \frac{\alpha_k^j}{\sqrt{2\omega_k(t)}}\;\exp\left(-i\,\int_{t_0}^t \omega_k(t') dt'\right) 
+ \frac{\beta_k^j}{\sqrt{2\omega_k(t)}}\;\exp\left(i\,\int_{t_0}^t \omega_k(t') dt'\right)
\ee
where the integral is taken over the intervals between $t_0$ and $t$ where $w_k^2>0$, and $\alpha_k^j$ and 
$\beta_k^j$ are constant coefficients (in the adiabatic approximation) during this time interval,
determined as we describe below. These correspond to the 
Bogoliubov coefficients (normalized as $|\alpha_k^j|^2-|\beta_k^j|^2 = 1$) and the initial vacuum for 
$t \rightarrow t_0$ is defined by the positive-frequency mode, $\alpha_k^0=1, \,  \beta_k^0=0 $.
The adiabatic invariant
\be
\label{number}
n_k^j = |\beta_k^j|^2
\ee
corresponds to the occupation number of the $\chi$-particles after $j$ inflaton 
oscillations and will be the major object of interest.

 For $t_{kj}^- < t < t_{kj}^+$ (below the barrier), the WKB
 approximation gives a superposition of
exponentially increasing and decreasing  solutions
\be
\label{chiOm}
\chi_k(t) \simeq \frac{a_k^j}{\sqrt{2\Omega_k(t)}}\;\exp\left(-\int_{t_{kj}^-}^t \Omega_k(t') dt'\right) 
+ \frac{b_k^j}{\sqrt{2\Omega_k(t)}}\;\exp\left(\int_{t_{kj}^-}^t \Omega_k(t') dt'\right)
\ee
where $a_k^j$ and $b_k^j$ are constant coefficients (normalized as $a_k^j\,b_k^{j\,*} - a_k^{j\,*}\,b_k^{j} = i$). 
Finally, for $t > t_{kj}^+$, we have $\chi_k(t) \simeq \chi_k^{j+1}(t)$ as given by (\ref{chiom})
 in the WKB regime with the shift $j \to j+1$ 

The solution (\ref{chiOm}) with non-vanishing $b_k^{j}$  corresponds to an exponentially fast prodution of particles. 
The WKB approximation in this tachyonic (below the barrier) stage is accurate for $A_k < 2q - 2\sqrt{q}$.
The duration of the tachyonic
 stage decreases with increasing $k$, that is with growing $A_k \,$. For $2q - 2\sqrt{q} < A_k < 2q$
 there is a short tachyonic stage, during which however the WKB approximation does not apply. 
As $A_k$ increases towards $2q$, the $\chi$-modes spend less and less 
time in the tachyonic regime, which becomes less and less efficient. For $A_k > 2q - 2\sqrt{q}$, particle 
production occurs in a small interval around $\phi = -\Phi$, providing a smooth limit with the
 case of broad parametric resonance for $A_k \ge 2q$. 

To evaluate the growth of $n_k$ during one oscillation of the inflaton, we have to match between 
the successive approximate solutions (\ref{chiom}) and (\ref{chiOm}) around the turning points, where 
the WKB approximation is inapplicable. This may be done by formally extending the domain of definition of 
$\chi_k$ to the complex plane, see Appendix A. Except for the normalization conditions, the procedure is 
similar to the calculation, in quantum mechanics, of the (spatial) evolution of a wave function in the quasi-classical 
regime \cite{LL3}. We then find the ``transfer matrix'' between the Bogoliubov coefficients after and before the 
$(j+1)^{\mathrm{th}}$ oscillation of the inflaton\footnote{This may be expressed as usual in terms of reflection and 
transmission coefficients satisfying $|R_k|^2+|D_k|^2 = 1$, up to exponentially small terms,
 $\propto {\rm exp } \left( - X_k^j \right) \,$, which are beyond the accuracy of the WKB approximation.}
\be
\label{transmat}
\left(\begin{array}{c}\alpha_k^{j+1}\vspace*{0.2 cm}\\\beta_k^{j+1}\end{array}\right) = 
e^{X_k^j}\;\left(\begin{array}{cc}
1  &  i\,e^{2i\theta_k^j} \vspace*{0.2 cm}\\ 
-i\,e^{-2i\theta_k^j} & 1 \end{array}\right)\;
\left(\begin{array}{c}\alpha_k^{j}\vspace*{0.2 cm}\\\beta_k^{j}\end{array}\right)
\ee
where 
\be
\label{Xkj}
X_k^j = \int_{t_{kj}^-}^{t_{kj}^+} \Omega_k(t')\,dt' 
\ee
and $\theta_k^j$ is the total phase accumulated from $t_0$ to $t_{kj}^-$ during the intervals where 
$\omega_k^2 > 0$, $\theta_k^j = \int_{t_0}^{t_{kj}^-} dt\,\omega_k(t)$. 

When the expansion of the universe is neglected, $\Phi$, $t_{k}^{\pm}$ and $X_k$ do not depend on $j$. 
We have furthermore in this case $\theta_k^j = \theta_k^0 + j\,\Theta_k$ where 
\be
\label{thetak}
\Theta_k = \int_{t_k^+}^{t_k^- + \frac{2\pi}{m}} \omega_k(t')\,dt'
\ee
is the phase accumulated during one inflaton oscillation when $\omega_k^2 > 0$. It is then easy to apply 
(\ref{transmat}) $(j+1)$ times recursively. With the initial conditions $\alpha_k^0 = 1$ and $\beta_k^0 = 0$, 
one finds the occupation number of the $\chi$-particles in the $k$-mode (with $k^2 < \sigma\,\Phi$) after $j$ oscillations of the inflaton to be
\be
\label{nkj0}
n_k^j = |\beta_k^j|^2 = \exp(2 j X_k) \; (2\,\cos \Theta_k)^{2(j-1)} \ .
\ee
This simple formula is the main analytic result of our paper.
 ${\rm exp } \left(2 \, X_k \right)$ gives the occupation number after the first oscillation. 
 For  the trilinear interaction (\ref{omexp}), one finds from (\ref{Xkj}) and (\ref{thetak}),
 in terms of the variables in the Mathieu equation (\ref{mathieu}) 
\be
\label{Xk}
X_k = \int_{\pi-\tilde{z}_k}^{\pi+\tilde{z}_k} \Omega_k(z)\; dz = 2\,\sqrt{2q-A_k} \; E\left(\tilde{z}_k \, ; \, 
\frac{4q}{2q-A_k}\right)
\ee
and 
\be
\label{tthetak}
\Theta_k = \int_{\tilde{z}_k}^{\pi-\tilde{z}_k} \omega_k(z)\; dz = 2\,\sqrt{2q+A_k} \; E\left(\frac{\pi}{2}-\tilde{z}_k 
\, ; \, \frac{4q}{2q+A_k}\right)
\ee
where $E(\theta \, ; \, m)$ denotes the incomplete elliptic integral of the second kind with amplitude 
$\theta < \frac{\pi}{2}$ and parameter $m$ \cite{abram}. Here we have defined
\be
\tilde{z}_k = \frac{1}{2}\,\arccos\left(\frac{A_k}{2q}\right) \in [0, \frac{\pi}{4}] 
\ee
so that the turning points are given by $z_k^{\pm} = \pi \pm \tilde{z}_k$.

The functional dependence of $X_k$, which controls the efficiency of particle production, is
\begin{equation}
X_k = 2 \sqrt{2 q} \:  f \left( \frac{A_k}{2 q} \right) \ .
\end{equation}
In the interval we are interested ($A_k < 2 q$), 
we found a good approximation  $f \left( y \right) \simeq 0.6 \left( 1 - y \right)$ (verified numerically).
 Therefore, we can use the following accurate approximation
\be
\label{appXk}
X_k \simeq - \frac{x}{\sqrt{q}}\,A_k + 2x\,\sqrt{q}
\ee
with $x=\frac{\sqrt{\pi}}{2\sqrt{2}} \frac{\Gamma(3/4)}{\Gamma(5/4)} \simeq 0.85$.

\begin{figure}[hbt]
\begin{center}
\begin{tabular}{cc}
\includegraphics{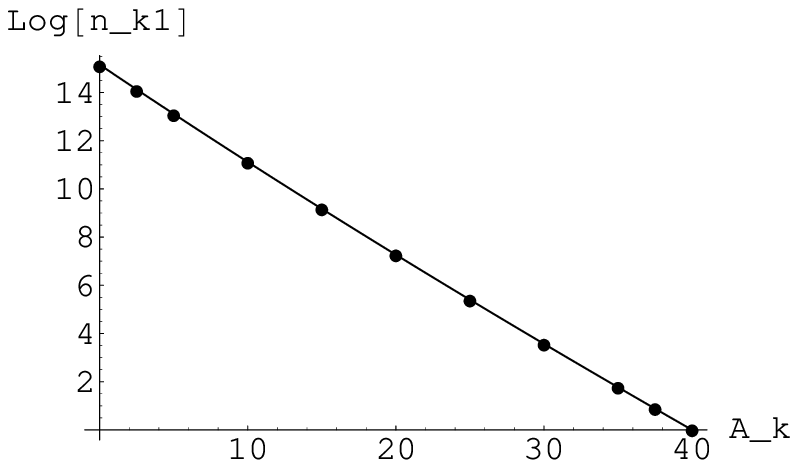}
\includegraphics{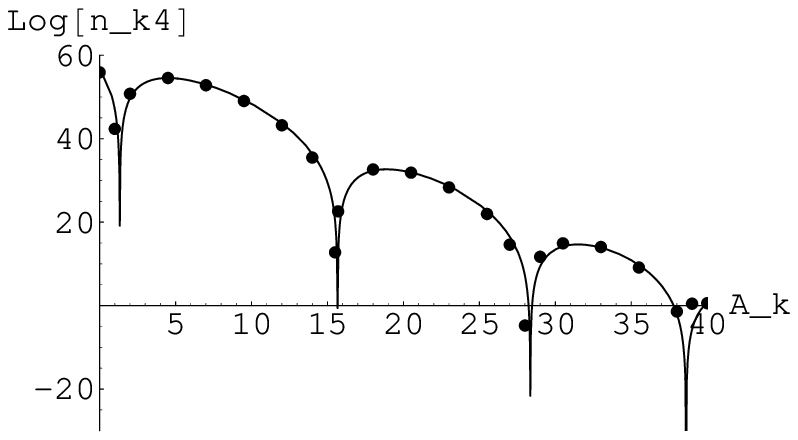}
\end{tabular}
\caption{$\log n_k^j$ derived from (\ref{nkj0}) as a function of $A_k$, after $j=1$ (left) and $j=4$ (right) 
oscillations of the inflaton, for $q=20$. The dots correspond to numerical solutions.}
\label{nkq20}
\end{center}
\end{figure}

The analytic formula (\ref{nkj0}) derived with the WKB method, gives a
pretty good approximation to the actual field dynamics.
To show this, we plot in 
Figure~\ref{nkq20} the occupation number $n_k^j$ in (\ref{nkj0}) as a function of $A_k$, after $j=1$ and $j=4$ oscillations, 
for the value of the parameter  $q=20$. 
Numerical calculations (dots) coincide with the results of the analytical curve.

Quite interestingly, after several inflaton oscillations ($j > 1$), there are
modes with negative frequency squared for which the tachyonic growth
does not occur (see right panel of Figure~\ref{nkq20}). These form
stability bands, which shrink to lines for $q >> 1$, given by $\cos
\Theta_k = 0$ (see (\ref{nkj0})). In some sense, the situation is
reversed with respect to the conventional parametric resonance for $q \ll 1$, where
only some values of $k$, forming the instability bands, ``resonate''
with the frequency of the inflaton and undergo an exponential
growth. Here, all the modes are amplified by the tachyonic instability
except the ones belonging to the very narrow stability bands. These are described in 
more details in Appendix B. Beyond the WKB approximation, each stability line 
is split after each inflaton oscillation. After several oscillations, the net effect  
is the formation of stability bands of finite width $\delta A_k \sim
\sqrt{q}\,e^{-x\sqrt{q}}$. As $q$ decreases, these bands become wider,
while the distance between them decreases as $\Delta A_k \sim
\sqrt{q}$ (see appendix B) ; at the same time the efficiency of particle production
decreases as $X_k \sim \sqrt{q}$ from (\ref{appXk})). When $q<1$, the
width of the stability and instability bands are comparable, and the
tachyonic effect becomes indistinguishable from (narrow) parametric
resonance (as is clear from the stability/instability chart of the
Mathieu equation in this region of parameters).

The expansion of the universe can be incorporated ``adiabatically'', which is to say that
the above formulas will stay the same but the parameters will depend
on $a(t)$. 
In an expanding universe,  we may essentially repeat the discussion 
leading to (\ref{transmat}) without any modification, because of the hierarchy of scales 
$\dot{\Phi}/\Phi \sim \sqrt{\Delta} \sim H << m << \sqrt{\sigma\Phi}$
at the beginning of preheating. The squares of the
physical momenta now redshift as $a^{-2}$ with the scale factor, which is faster than the amplitude of the 
inflaton oscillations, $\Phi \sim a^{-3/2}$. This means that more and more modes satisfy $k^2/a^2 < \sigma\,\Phi(t)$ 
and are therefore amplified by the tachyonic instability. The modes
which are amplified latest give the highest 
contribution to the comoving phase space volume. However, this effect is subdominant with respect to the decrease 
with time of the rate of particle production $X_k$. Using (\ref{appXk}), $X_k^j$ during the $j^{\mathrm{th}}$ inflaton 
oscillation may be estimated as 
\be
\label{appXexp}
X_k^j \simeq  - \frac{x\,A_k}{\sqrt{q_0}\,a_j^{5/4}} + 2x\,\frac{\sqrt{q_0}}{a_j^{3/4}} 
\ee
where $q_0$ is the initial $q$-parameter and $a_j$ is the mean value of the scale factor during the $j^{\mathrm{th}}$ 
period of particle production. Since the second term dominates this expression, $X_k^j$ decreases approximately as 
$X_k^j \sim \sqrt{q} \sim a_j^{-3/4}$ with the expansion. 
As in the case of broad parametric resonance, the phase $\theta_k^j$ in (\ref{transmat}) is random in an 
expanding universe. This results here in broader ``stability bands'' as compared to flat spacetime. Contrary to preheating 
due to the $\phi^2\,\chi^2$ interaction, the interference term in $\theta_k^j$ in (\ref{transmat}) is never large enough to 
decrease the total number of particles during some oscillations of the inflaton. In conclusion, the main effect of the expansion 
of the universe is a decrease of the efficiency of particle production at each inflaton oscillation, due to the decrease of the 
effective $q$-parameter. This tends to lengthen the preheating stage. If the $q$-parameter decreases up to $q \sim 1$, we
 are then 
in the regime of narrow parametric resonance, which is shut off by the expansion of the universe 
as discussed in \cite{KLS97}. In general, however, the actual dynamics at the end of preheating is rather 
complicate and includes backreaction of the quanta produced, rescattering and self-interaction.

The analytic derivation leading to the formula (\ref{nkj0}) is valid for tachyonic values of $\omega^2_k$, which can include
$\phi\chi^2$ and $\phi^2\chi^2$ interactions, as far as the three-legs coupling dominates.
The tachyonic effect is stronger for stronger three-legs couplings (relatively to the four-legs couplings).
In the limiting case, when the $\phi^2\chi^2$ interaction is absent,
 preheating is so efficient that it completes typically after the 
first or second oscillation of the inflaton homogeneous mode. Indeed, as the $\chi$-modes are amplified, 
their variance $\langle\chi^2\rangle$ grows, and the contribution of the $\lambda\,\chi^4$ self-interaction to their 
frequency squared, $3\lambda\,\langle\chi^2\rangle$, eventually becomes significant. The tachyonic instability 
is then shut off  when $\langle\chi^2\rangle \simeq \sigma\,\Phi(t)/(3\lambda)$. (In principle, some 
modes may still be amplified by parametric resonance during the next inflaton oscillation.) Let us roughly 
estimate the value of the $q$-parameter for which this occurs after the first oscillation of the inflaton. 
For this me may neglect the expansion of the universe and work with the value of $q$ during the first 
``tachyonic interval'', which is about one half of its initial value $q_0$ at the end of inflation. From (\ref{appXk}),
 the variance of $\chi$ after the first oscillation of the inflaton may 
be estimated as
\be
<\chi^2> \sim \frac{1}{2\pi^2}\;\int_0^{+\infty} dk\,k^2\,|\chi_k^1|^2 \sim 
\frac{1}{2\pi^2}\;\int_0^{\sqrt{\sigma\,\Phi}} dk\,k\,e^{2X_k} \sim \frac{m^2\,\sqrt{q}\,e^{4x\sqrt{q}}}{32\,x\,\pi^2} 
\ee
For instance for  the  value $\lambda = \sigma^2/m^2$, this is equal to $\sigma\,\Phi/(3\lambda)$ for $q \sim 50$ 
and $q_0 = 100$. 
The corresponding energy density in $\chi$-particles may be estimated in the same way
\be
\rho_{\chi} \sim \frac{1}{2\pi^2}\;\int_0^{+\infty} dk k^2\,\omega_k\,n_k^1 \sim 
\frac{1}{2\pi^2}\;\int_0^{\sqrt{\sigma\,\Phi}} dk\,k^3\,e^{2X_k} \sim \frac{m^4\,q\,e^{4x\sqrt{q}}}{256\,x^2\,\pi^2}
\ee
which is about $5$ percent of the initial energy density of the system, $\rho_i \sim \frac{1}{2}\,m^2\,\Phi_i^2$. 
These estimations are over-simplified but the orders of magnitude are correct. In particular, most of the energy 
density is still in the coherent inflaton mode at the end of preheating, and it will mainly decay during the 
``rescattering'' stage.

In this section we outlined  in detail the formalism of particle
creation from a $\sigma \phi \chi^2$ interaction.
The formalism in fact is more general, and can be applied to the
more general case when both $\sigma \phi \chi^2$ and $g^2 \phi^2 \chi^2$ interactions contribute.
Instead of repeating the procedure for this case, in the next section we will
discuss when each of them may dominate.

\section{Comparison of the Effects of Three and Four Legs Interactions}\label{sec:comparison}

In this section we will address the important question: 
Which interaction,  three- or four-legs, is dominant during preheating ? 
We shall estimate the parameter ranges for which parametric resonance or tachyonic 
amplification will dominate at the stage of preheating, and discuss how the
three-legs interaction enhances parametric resonance even if the
trilinear term is subdominant. 

After inflation the inflaton oscillates sinusoidally with a decaying
amplitude $\Phi(t) =\Phi_0/a^{3/2}$. When the backreaction of the $\chi^4$ 
self-interaction in (\ref{v4}) is still negligible, the time-dependant 
frequency of the $\chi$-modes is given by
\begin{equation}
\label{w34}
\omega_k^2(t) = k^2 + \sigma\,\Phi(t)\,\sin(mt) + g^2\,\Phi(t)^2\,\sin^2(mt) \ ,
\end{equation}
where $k$ denotes the physical momenta. Without expansion of the universe, 
this reduces to a particular case of the Hill equation  
\be
\chi_k '' + \left( A_k - q_3 \cos z +  \frac{q_4}{2} \, \cos 2 \, z \right) \chi_k = 0 \ ,
\label{hill}
\ee
where its parameters are slightly different from those in (\ref{mathieu}), namely
$z=mt + \pi /2$, $A_k =k^2 / m^2 + q_4/2$ and we have defined
the dimensionless quantities
\be
\label{q34} 
q_3 = \frac{\sigma\,\Phi}{m^2} \;\;\;\; \mbox{ and } \;\;\;\; q_4 = \frac{g^2\,\Phi^2}{m^2} \ .
\ee

Equation (\ref{hill}) is an equation with periodic coefficients, therefore, according to 
the Floquet theorem, 
it admits  exponentially unstable solutions.
However, the structure of the stability/instability chart will be denser and  more complicated than that of the 
Mathieu equation. In any case the expansion of the universe for the interesting case of large parameters $q_3$ and $q_4$
makes the stability/instability chart relevant only heuristically.
Therefore we restrict ourself to qualitative arguments based on the physical picture
of particle creation from three- and four-legs vertices. 

First  consider  the 
case $q_3 \leq q_4$, for which the frequency squared (\ref{w34}) is negative in the interval 
$-q_3/q_4 < \sin(mt) < 0$. This leads to a tachyonic amplification  of modes with 
$\frac{k^2}{m^2} < \frac{q_3^2}{4q_4}$. In addition to this effect,
modes are produced by the 
four-legs interaction
 when the non adiabaticity condition, $|\dot{\omega}_k| > |\omega_k|^2$, is satisfied. 
This happens in small intervals $\Delta \sin(mt) \sim 2/q_4^{1/4}$ around $\sin(mt) = -q_3/2q_4$, 
as may be found by applying the calculation of \cite{KLS97}. The typical momentum of the modes 
amplified by this process is given by $\frac{k^2}{m^2} = \frac{q_3^2}{4q_4} + \frac{\sqrt{q_4}}{2}$. 
Inspecting the right hand side of this expression, we can break the range of $q_3$ and $q_4$ into two regions,
  $q_3 < q_4^{3/4}$ and $q_3 > q_4^{3/4}$.

For the case
\be
\label{q4}
 q_3 < q_4^{3/4}
\ee
the periods of tachyonic instability are negligible, and particle production 
 occurs due to the non-adiabaticity  in  very small intervals $m\,\Delta t << 1$
due to the four-legs vertex. Yet, the presence of a three-legs vertex facilitates the effect. 
Indeed, in this case, 
we may use the method of successive parabolic scatterings  \cite{KLS97} 
to evaluate the growth of the occupation number during one event of particle production.
After tedious calculations we find 
\be
\label{kls}
n_{j+1} = n_j + 
2\,e^{-\pi\kappa^2}\;\left(1-\sin\hat{\theta}\,\sqrt{e^{\pi\kappa^2} +
1}\right) n_j
\ee
where $\hat{\theta}$ is the phase of waves accumulated between successive scatterings.
In an expanding universe and for large $q$ parameters, $\hat{\theta}$ can be treated as the
random phase so that the resonant effect is stochastic.
(In a sense the stability/instability bands are smeared, see \cite{KLS97} for details.)
In formula (\ref{kls})
\be
\label{klsk}
\kappa^2 = \frac{k^2}{m^2}\,q_4^{-1/2} - \frac{q_3^2}{4q_4^{3/2}} \ ,
\ee
which is different from $\kappa^2 = \frac{k^2}{m^2}\,q_4^{-1/2}$ for
the case when only the four legs interaction is 
present. Thus, for the range of parameters (\ref{q4})
when  preheating is  dominated by parametric resonance of the four-legs interaction, 
 the trilinear interaction enhances  particle production
due to the additional 
factor $e^{\pi q_3^2/4q_4}$ in (\ref{kls}).

In  the opposite case 
\be
\label{q3}
 q_3 > q_4^{3/4}
\ee
the tachyonic effect is dominant, and we have to use formula (\ref{nkj0}) of 
the previous section.

We are especially interested in the SUSY case. Let us take the superpotential (\ref{w4}), which leads to  
 $q_3 = \sqrt{q_4}$. For $q_3, q_4 \gg 1$ this corresponds to the case (\ref{q4}). 
This means that
  in the SUSY theory  where both three and four legs interactions are present,
the stage of preheating is dominated by the four legs interaction.

Suppose that initially we have condition (\ref{q4}), as in the SUSY theory.
Consider $q_4^{3/4}/q_3$ as function of time.
The inflaton amplitude redshifts as $\Phi
\sim (mt)^{-1}$ with the expansion of the universe, so that the
critical ratio $q_4^{3/4}/q_4$ decreases with time as $(mt)^{-1/2}$.
If initially this ratio is larger than unity, it decreases with time
and later on the situation reverses,  $q_4^{3/4}/q_4$ becomes less than unity, 
and the three legs vertex dominates.

We conclude that in many cases, notably in the SUSY theory,
the stage of preheating is dominated by the effects of the four legs interaction,
while the later stages are dominated by the three legs interaction. As we will see in the next 
section, independently of the dominant mechanism for preheating, the trilinear interaction 
dominates the following turbulent dynamics.

\section{Non-linear Dynamics After Preheating with a  Three Legs Interaction}
\label{numsimul}

So far we have analytically studied preheating with a three-legs interaction
and compared the effects of three- and four-legs interactions during the
linear stage. After the system becomes non-linear, the analytic theory
is replaced by numerical simulations of the dynamics.
In this section we report the results of those simulations. 

We will consider three different models, one
with only a four-legs interaction, one with only a three-legs interaction, and one
with both three- and four-legs interactions.
Let us define these three models as
\begin{eqnarray}
\label{models}
&\mbox{Model A: }& \;\;\;\;\; q_{4i} = 10^4 \;\;\;\; , \;\;\;\; q_{\chi i} = 5\,\times 10^3 \;\;\;\; , \;\;\;\; q_{3i} = 0 \\
&\mbox{Model B: }& \;\;\;\;\; q_{4i} = 10^4 \;\;\;\; , \;\;\;\; q_{\chi i} = 5\,\times 10^3 \;\;\;\; , \;\;\;\; q_{3i} = 10^2 \\
&\mbox{Model C: }& \;\;\;\;\; q_{4i} = 0 \;\;\;\;\;\;\;\; , \;\;\;\; q_{\chi i} = 10^4 \;\;\;\;\;\;\;\;\;\; , 
\;\;\;\; q_{3i} = 10^2 \\
\end{eqnarray}
where we have introduced the dimensionless quantities $q_{4i} =
g^2\Phi_0^2/m^2$, $q_{3i} = \sigma\Phi_0/m^2$  and $\Phi_0 = 0.193\,M_P$ is the initial
amplitude of the inflaton zero-mode at the end of inflation.
The parameter $q_{\chi i} =\lambda\Phi_0^2/m^2$ characterizes the self-interaction of the $\chi$-field.

 Model A
is a standard preheating model with parametric resonance due to a four-legs interaction,
without three-legs coupling.
 Model B contains both three- and four-legs couplings and 
corresponds to the SUSY case discussed in Section~\ref{3pot} with
$\lambda = g^2/2$ and $\sigma = g\,m$ in (\ref{v4}). 
Model C corresponds to the potential (\ref{v3}) without  
$\phi^2\,\chi^2$ interaction (when the exact flat direction shown in Figure~\ref{vplot} is slightly lifted, 
$\lambda > \sigma^2/(2m^2)$) and
thus allows us to see the pure effects of tachyonic resonance
due to the three-legs coupling.  A $\chi^4$ self-interaction is present in all models (to ensure the stability 
of the potential in the presence of the $\phi\,\chi^2$ interaction, see Section~\ref{3pot}). We will see 
that this self-interaction does not change qualitatively the dynamics.

We used the LATTICEEASY program \cite{latticeeasy} to perform
numerical lattice simulations for the theory (\ref{v4}) in an
expanding universe.
For the three models, we used a 3-dimensional lattice with $N=256^3$
points and a comoving edge size $L=10/m$.  This allows us to probe
comoving momenta in the range $0.6\,m < k < 140\,m$. Here we are
interested in the behavior of the system both during and after preheating.
 Long time integrations require both a high ultra-violet (UV) cutoff for the modes 
and a high resolution of modes in the 
infra-red (IR) part of the spectra. This motivates the
value of $q_{4i}$ chosen, which is large enough to produce efficient
preheating, but also small enough to produce quick rescattering
\cite{podol}. For the three models, the simulations were performed up
to a total period of $m\,\Delta t = 1000$ starting from the end of
inflation.  We also checked the robustness of
our results by performing 2-dimensional simulations with both a larger
UV cutoff (smaller lattice spacing) and a higher resolution in the IR
(larger box size).

As output of the calculations, we present the evolution of the mean
values of the fields, the
evolution of the occupation number spectra, and most importantly the evolution
of the equation of state. For the definitions of these quantities, see 
e.g.~\cite{podol}. As we will see, after the preheating stage
model B (with both types of interactions) will behave qualitatively
similarly to model C (with only a three legs interaction), but significantly
differently from model A (with only a four legs interaction). We can
thus conclude that the three legs interaction dominates the dynamics
of model B during this stage.

\begin{figure}[hbt]
\begin{center}
\begin{tabular}{ccc}
\includegraphics[width=6cm]{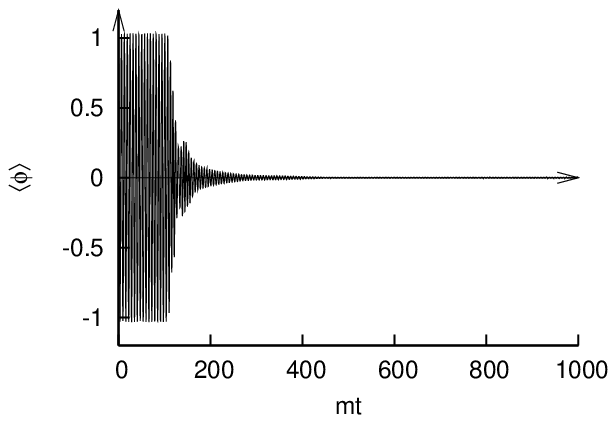}
\includegraphics[width=6cm]{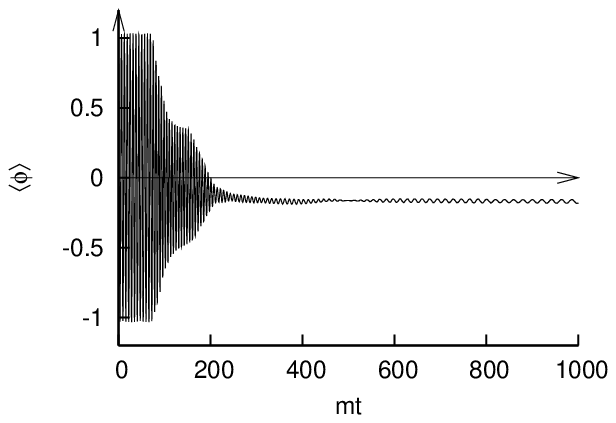}
\includegraphics[width=6cm]{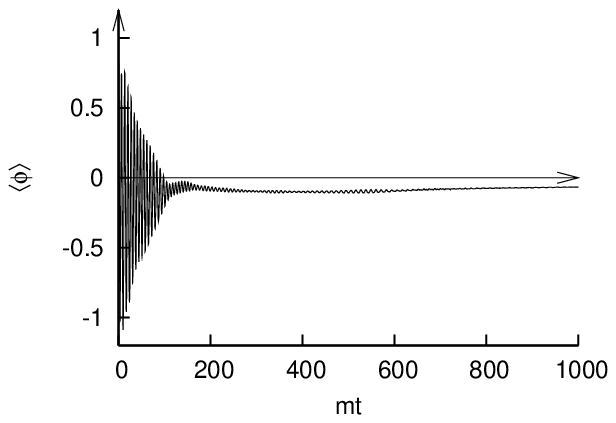}
\end{tabular}
\caption{Mean value of $\phi$, in units of $\Phi_0$ and multiplied by
$a^{3/2}(t)$, as a function of $m\,t$, for models A (left), B (middle),
and C (right).}
\label{means0}
\end{center}
\end{figure}

The evolution with time of
the homogeneous inflaton zero-mode in the 3 models is shown in Figure~\ref{means0}. 
In model A without the trilinear term the inflaton mean decays to zero, but in the
presence of the trilinear term in models B and C the backreaction of the quanta produced 
displaces the global minimum of the potential away from $\phi = \chi = 0$.
Therefore $\phi$ develops a non vanishing VEV $\bar{\phi} = \phi - \delta\phi$, which is 
in very good agreement with the one found from the Hartree approximation 
$\bar{\phi} \simeq {-\,\frac{\sigma}{2}\,\langle\chi^2\rangle}/({m^2 + g^2\,\langle\chi^2\rangle})$
where $\langle\chi^2\rangle$ is the variance of $\chi$.

\begin{figure}[hbt]
\begin{center}
\begin{tabular}{cc}
\includegraphics[width=6cm]{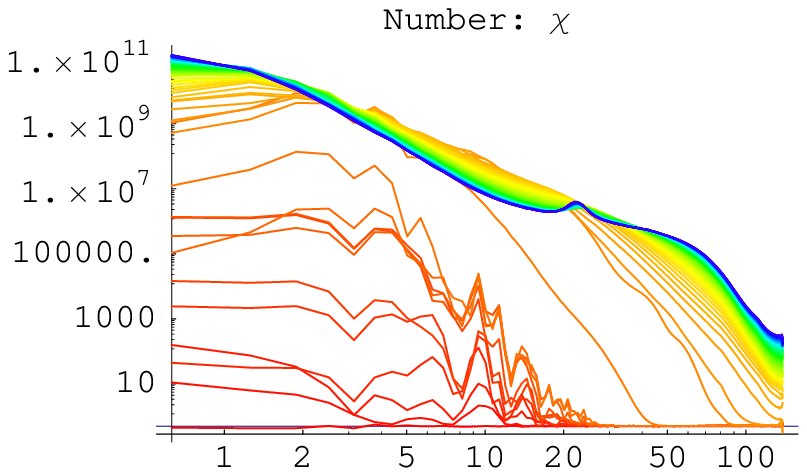}
\includegraphics[width=6cm]{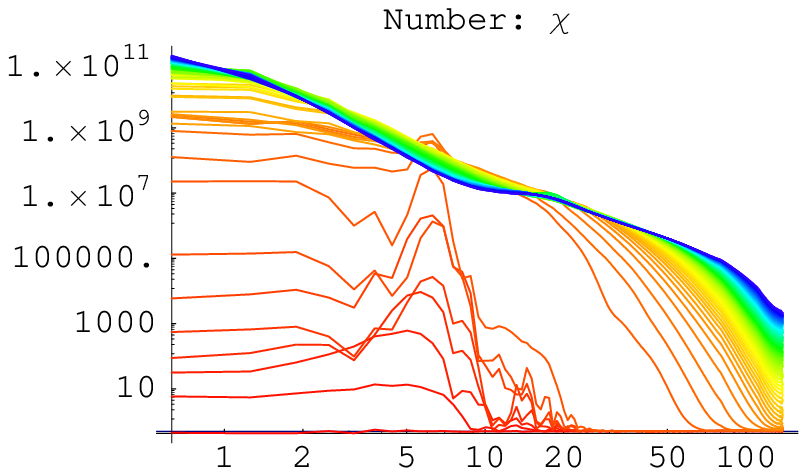}
\includegraphics[width=6cm]{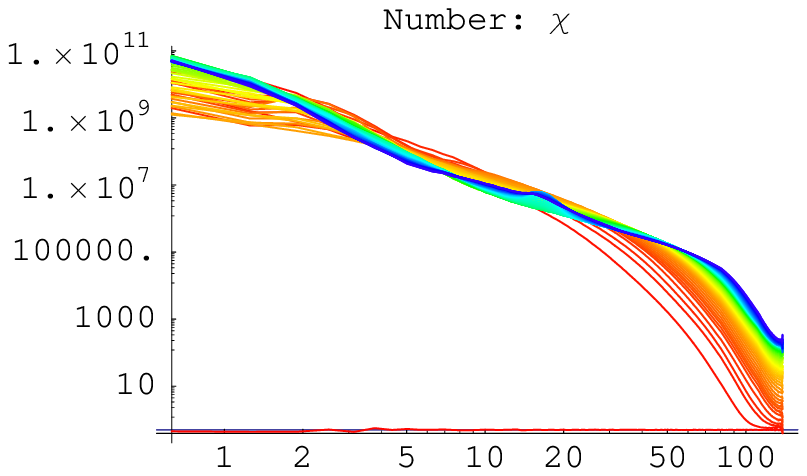}
\end{tabular}
\caption{Occupation number $n_k^{\chi}$ as a function of $k/m$ for models A (left), B (middle), and C
(right). The spectra go from early times (red plots) to late times
(blue plots) with spacing $\Delta t =
10/m$.}
\label{nk1}
\end{center}
\end{figure}

\begin{figure}[hbt]
\begin{center}
\begin{tabular}{cc}
\includegraphics[width=6cm]{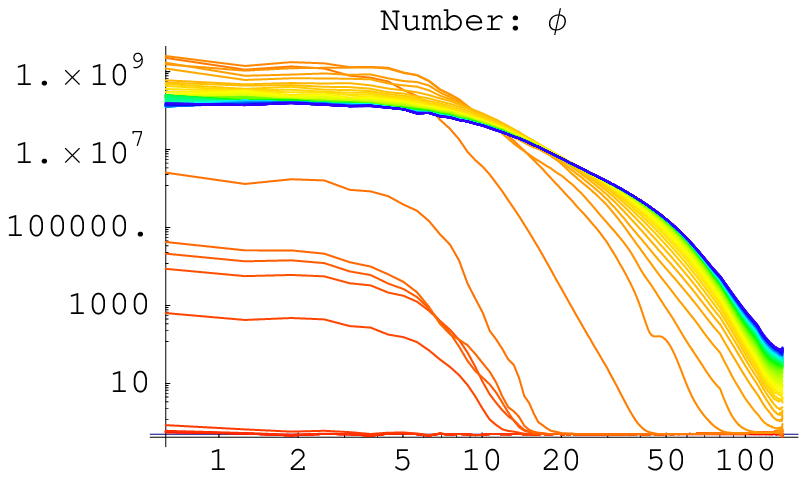}
\includegraphics[width=6cm]{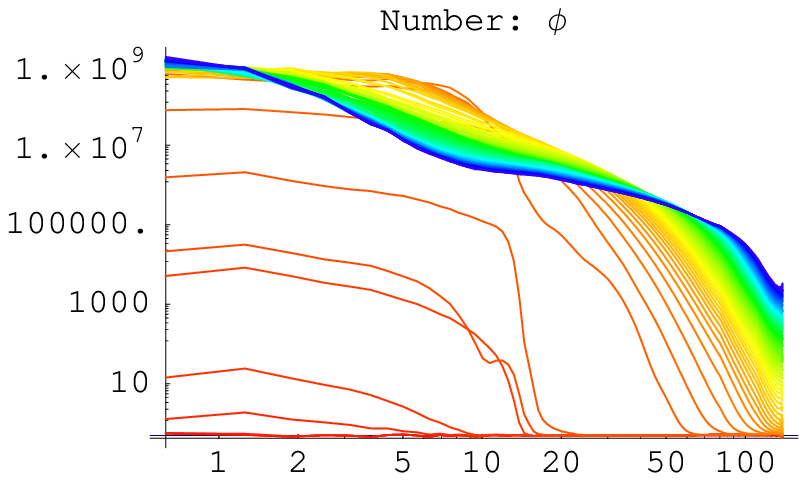}
\includegraphics[width=6cm]{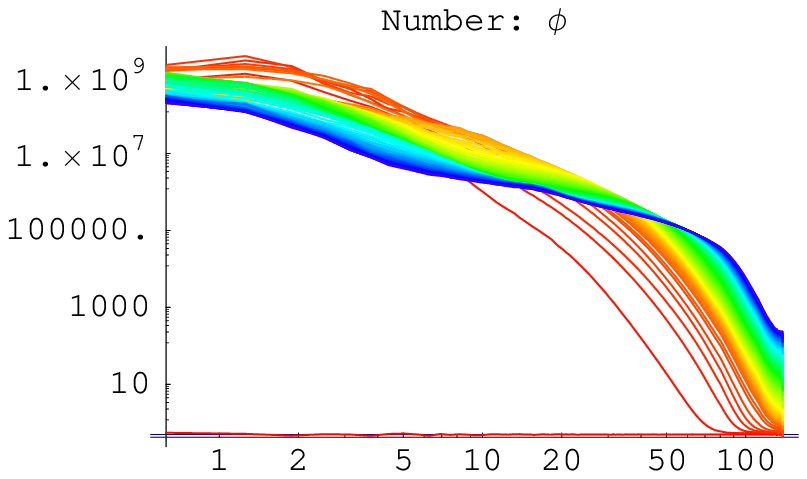}
\end{tabular}
\caption{The same as in Figure~\ref{nk1} but for $\delta \phi$ quanta.}
\label{nk0}
\end{center}
\end{figure}

The occupation numbers $n_k$ of $\chi$ and $\delta\phi$ quanta are
shown in Figures \ref{nk1} and \ref{nk0} for the three models.
We can distinguish three qualitatively different stages: linear preheating,
 the short violent non-linear stage and the long turbulent  stage 
where spectra cascade towards saturated distributions.
 The
preheating stage in which the spectrum of $\chi$ grows rapidly occurs
faster in model B than in model A, 
showing the effect of the trilinear
term on parametric resonance.
The theory of this effect (impact of three-legs vertices on parametric resonance)
was given in Section IV.
Meanwhile in model C, where preheating
occurs through tachyonic resonance only, this stage takes only one inflaton
oscillation, showing the efficiency of the effect.

\begin{figure}[hbt]
\begin{center}
\includegraphics[width=7cm]{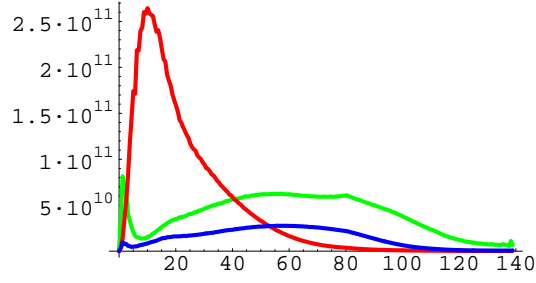}
\caption{Spectra $k^2\,\omega_k\,n_k/m^3$ as a function of $k/m$ for
$\delta \phi$ quanta at $m\,t = 1000$. The spectrum peaked in the IR
is for model A while the spectra for model B (above) and model C (lower) 
are spread out much further in the UV.}
\label{k2wnk}
\end{center}
\end{figure}

After preheating ends there is a brief violent stage of rescattering,
followed by a much longer stage of weak turbulence during which the
spectra cascade towards UV and IR modes, gradually approaching
kinetic equilibrium. We can see in Figures \ref{nk1} and \ref{nk0} that this
cascade occurs faster in the presence of a trilinear term, i.e. in
models B and C compared to model A. In Figure~\ref{k2wnk} we plot $k^2
\omega_k n_k$ for $\delta \phi$ quanta, which corresponds to the
 contribution of these modes to the total energy density of the
system at the end of the simulations. We see that in the presence of
the trilinear interaction the spectrum is spread much more strongly towards
the UV. This means that more $\delta \phi$ quanta are relativistic
in this case. This will be important for the following discussion.

\begin{figure}[hbt]
\begin{center}
\begin{tabular}{cc}
\includegraphics[width=6cm]{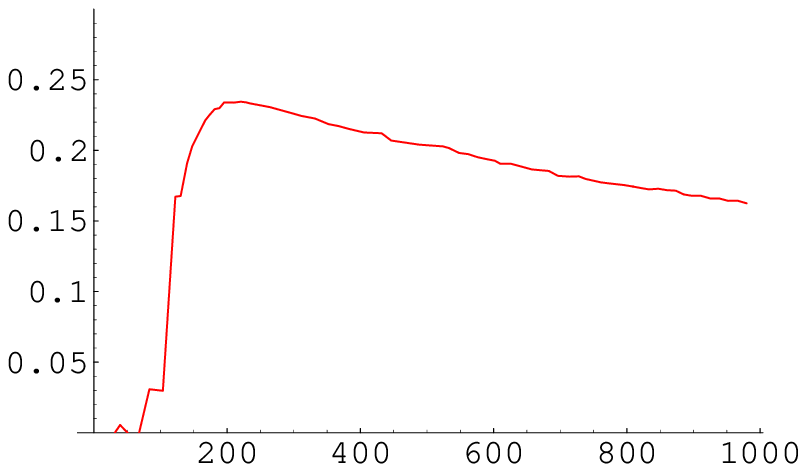}
\includegraphics[width=6cm]{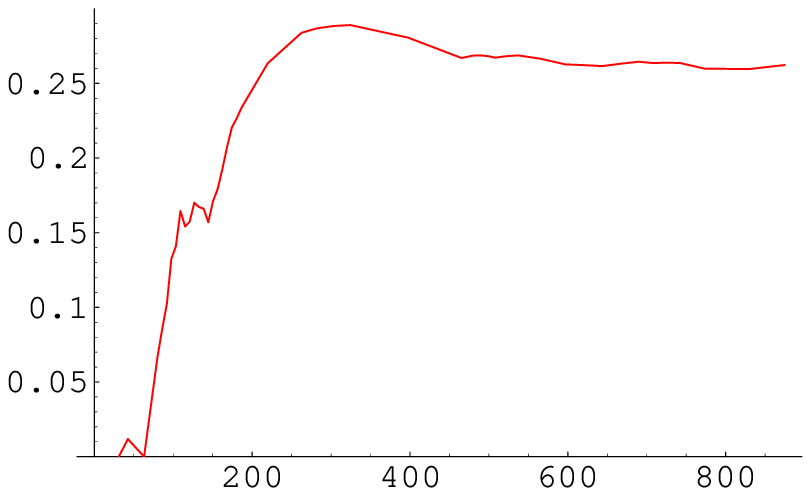}
\includegraphics[width=6cm]{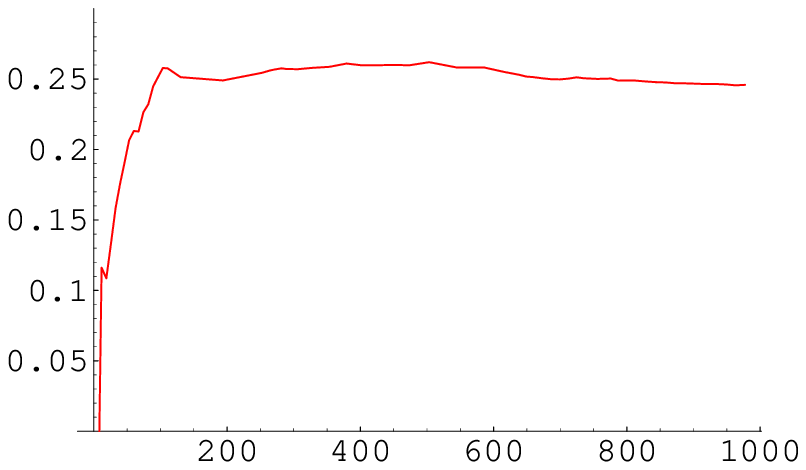}
\end{tabular}
\caption{Equation of state $w = p_{\mathrm{tot}}/\rho_{\mathrm{tot}}$
of the system, as a function of $m\,t$, for models A (left), B
(middle), and C (right).}
\label{eos}
\end{center}
\end{figure}

Now we turn to the evolution of the equation of state (EOS), see  Figure~\ref{eos}.
 In model A, where the trilinear interaction is absent, the equation of state
rises to roughly $1/4$ and then starts falling back towards zero. 
 In this model the EOS will never reach $w =1/3$ because the 
massive inflaton particles cannot decay completely \cite{podol}.
In the presence of a trilinear term, by contrast, the
equation of state jumps to a plateau
value and does not decrease.  In our simulations of classical fields 
this value is slightly below $1/3$. This is the most important reason to introduce a three-legs interactions.
The trilinear term dramatically  affects the evolution of the equation of
state.
It is expected that much later in the evolution, when quantum effects
 become important, the decay of the massive inflaton
will result in an
asymptotic radiation EOS.

\begin{figure}[hbt]
\begin{center}
\begin{tabular}{cc}
\includegraphics[width=6cm]{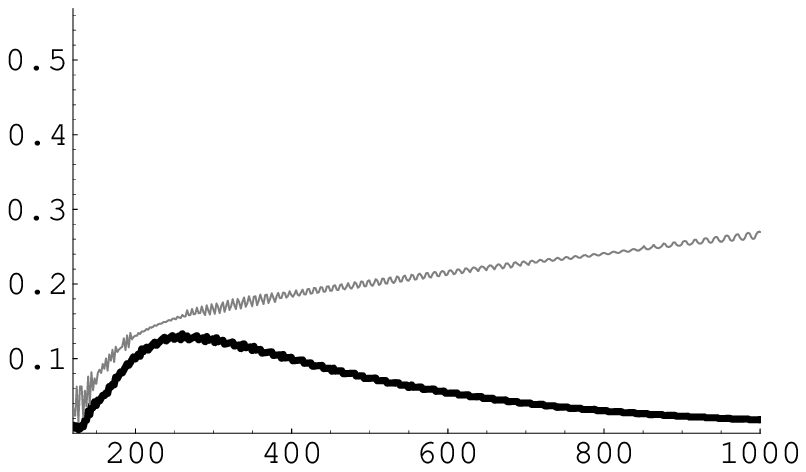}
\includegraphics[width=6cm]{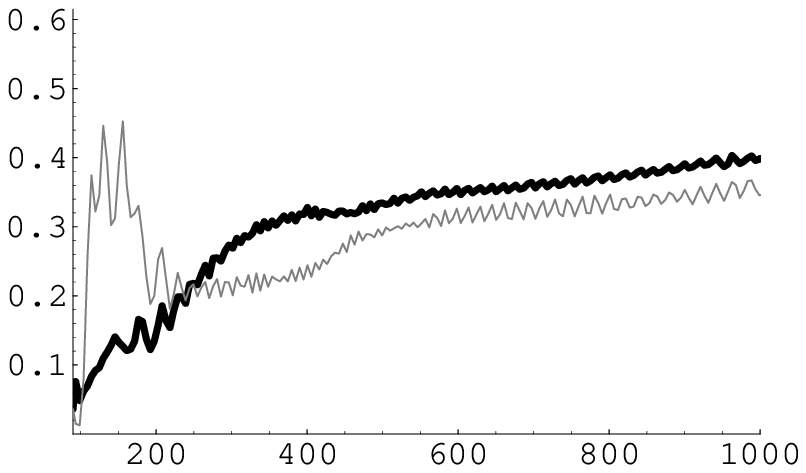}
\includegraphics[width=6cm]{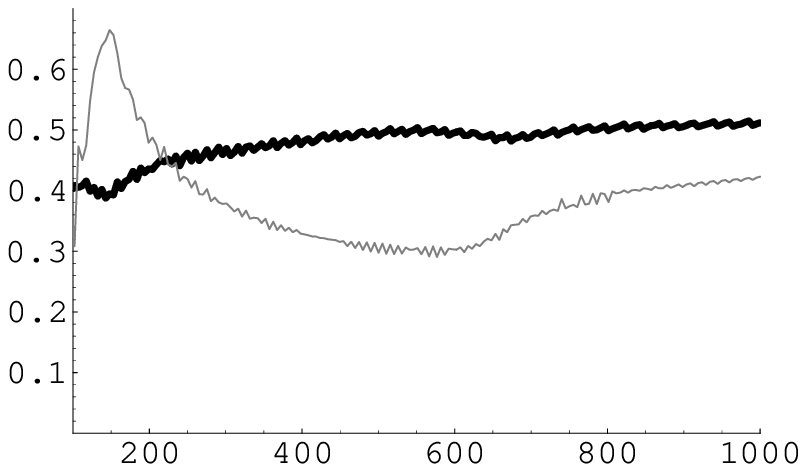}
\end{tabular}
\caption{The fraction of quanta that are relativistic,
$\frac{n_{\mathrm{rel}}}{n}$ as a function of $m\,t$
for $\delta\phi$ (dark) and $\chi$ (grey) in models A (left), B
(middle), and C (right).}
\label{nrel}
\end{center}
\end{figure}

To understand the behavior of the EOS, we investigate how the fraction of relativistic
particles is evolving in the system.
Let us consider the 
 fraction of
quanta whose physical momenta $k/a$ are larger than their effective
masses
\begin{equation}
\frac{n_{\mathrm{rel}}}{n} = \frac{\int_{k > a\,m_{\mathrm{eff}}}
d^3\mathbf{k}\,n_k}{\int d^3\mathbf{k}\,n_k} \ .
\label{relat}
\end{equation}
The effective masses are given by 
\begin{eqnarray}
m_{\phi, \mathrm{eff}}^2(t) &=& m^2 + g^2\,\langle\chi^2\rangle \\
m_{\chi, \mathrm{eff}}^2(t) &=& g^2\,\bar{\phi}^2 + \sigma\,\bar{\phi} + g^2\,\langle\delta\phi^2\rangle 
+ 3\lambda\,\langle\chi^2\rangle
\end{eqnarray}
for $\phi$ and $\chi$ respectively.
We calculate and 
plot in Figure~\ref{nrel} the fraction of relativistic particles.

After the
preheating stage, the effective mass of the $\chi$-quanta is dominated by
the term $3\lambda\,\langle\chi^2\rangle$, which is diluted with the
expansion of the universe as, approximatively, $m_{\chi, \mathrm{eff}}
\sim \sqrt{\langle\chi^2\rangle} \sim a^{-1}(t)$. This is the same
rate as the redshift of physical momenta, so that the number of
relativistic $\chi$-quanta increases due to the flow of comoving
momenta towards UV modes. On the other hand, the effective mass of
the $\delta\phi$-quanta is dominated by the constant bare mass
$m$. Therefore, these quanta become less relativistic as their
physical momenta are redshifted with the expansion, except if the flow
of comoving momenta towards UV modes is as fast as $a(t)$.  We see
from Figure~\ref{nrel} that this only occurs in the presence of a
trilinear term. In short, model A moves towards matter domination
because the $\phi$ particles cannot decay completely and become
increasingly non-relativistic. In the presence of a trilinear
interaction, however, the $\phi$ particles become increasingly
relativistic.
The inflaton field should decay completely on a much longer time scale through the
quantum process $\phi \to \chi \chi$.
 When the occupation numbers have been diluted by the expansion of the universe,
$n_k/a^3 << 1$, this process should dominate the dynamics of model B and C. In this regime,
 the typical time of inflaton
particle decay is given by $\tau \sim 1/\Gamma \sim 8\pi m/\sigma^2 \sim 10^8/m$ for $q_3 = 100$. 
This stage of the evolution cannot be captured by simulations of the
classical fields dynamics, however.

\section{Preheating with Non-Renormalizable Interactions}
\label{nonrenorm}

In this section we study preheating due to the presence of
non-renormalizable terms in the theory
\be
\label{nonr}
\Delta {\cal L}=-\lambda_n \frac{\phi^n \chi^2}{M^{n-2}} \ .
\ee
These terms may occur alongside the
usual three- and four-legs vertices, or in cases where the only
interactions are the
non-renomalizable ones.
We show that, quite surprisingly, the
non-renormalizable terms (\ref{nonr}) alone
lead to very efficient preheating.
We also show that, even in the presence of 
renormalizable interactions, non-renormalizable ones 
tend to dominate the early stage of preheating.
 
If $M$ is close to the Planck mass, the amplitude of the inflaton 
satisfies $\Phi/M < 1$ during preheating and the theory is in a 
controllable regime. If all the couplings $\lambda_n$ are of order 
one, the dominant interaction in (\ref{nonr}) is then the one with 
$n = 3$. Here we will consider the more general case where $n$ is odd. 
For $n$ even, there is no tachyonic amplification and preheating 
occurs due to parametric resonance. 

Let us begin with the theory of $\chi$ particle creation from the oscillating inflaton
due to the interaction  (\ref{nonr}) with $n$ odd,
without expansion of the universe.
The $\chi_k$ eigenmodes  obey the equation
\be \chi_k '' + \lmk A_k - 2 q_n \cos^n 2z \rmk \chi_k = 0 ,
\label{nmathieu}
\ee
where $mt = 2z - \frac{\pi}{2}$, $A_k = \frac{4k^2}{m^2}$ and $q_n =
\frac{2\lambda_n \Phi^n}{m^2 M^{n-2}}$. For $M \sim M_P$, $\lambda_n \sim {\cal O}(1)$ 
and $m \sim 10^{13}$ GeV, we have $q_n \sim 10^{12}/10^n$ at the beginning of preheating.

The method of solving this equation for $q_n \gg1$
will be based on a generalization of the
method of Section~\ref{tachres} which we developed for
equation (\ref{mathieu}) for the regime of tachyonic resonance.

Repeating the analysis of section~\ref{tachres} and the Appendices,
we find that the behavior of occupation numbers of $\chi$
particles with time is given by the formula which generalizes (\ref{nkj0})
\be
\label{nkjg}
n^{(n)}_j(k) = \exp(2 j X^{(n)}_k) \; \left(2\,\cos \Theta^{(n)}_k \right)^{2(j-1)} \ .
\ee
 with
\begin{equation}
X^{(n)}_k = \int_{\pi-\tilde{z}_{k,n}}^{\pi+\tilde{z}_{k,n}} \sqrt{2
q_n \cos^n 2z - A_k} \; dz
\label{Xkn}
\end{equation}
and
\be
\Theta^{(n)}_k = \int_{\tilde{z}_{k,n}}^{\pi-\tilde{z}_{k,n}} \sqrt{A_k - 2 q_n \cos^n 2z} \; dz ,
\label{Thetakn}
\ee
where $\tilde{z}_{k,n} = \frac{1}{2} \textrm{arccos} \lmk \lmk \frac{A_k}{2q_n} \rmk^{1/n} \rmk$. 
The condition $\cos \Theta^{(n)}_k=0$ defines the median locations of the stability bands.

To estimate the rate of particle production, we shall estimate the function
 $X^{(n)}_k$ given by the integral  (\ref{Xkn}). It turns out to be well approximated by the formula
\be
X_k \simeq \sqrt{2q_n} \; \frac{\sqrt{\pi}}{2} \frac{\Gamma \lmk \frac{2+n}{4} \rmk}{\Gamma \lmk 1+\frac{n}{4} \rmk} \; 
\lmk 1 - \lmk \frac{A_k}{2q_n}\rmk^{\frac{1+n}{2n}} \rmk \ .
\label{Xknap}
\ee
This reduces to (\ref{appXk}) for particular case of $n=1$.

\begin{figure}[hbt]
\begin{center}
\begin{tabular}{cc}
\includegraphics{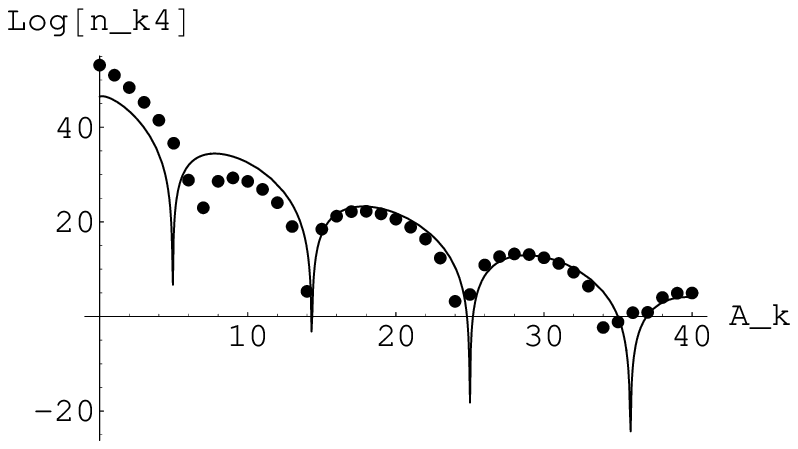}
\includegraphics{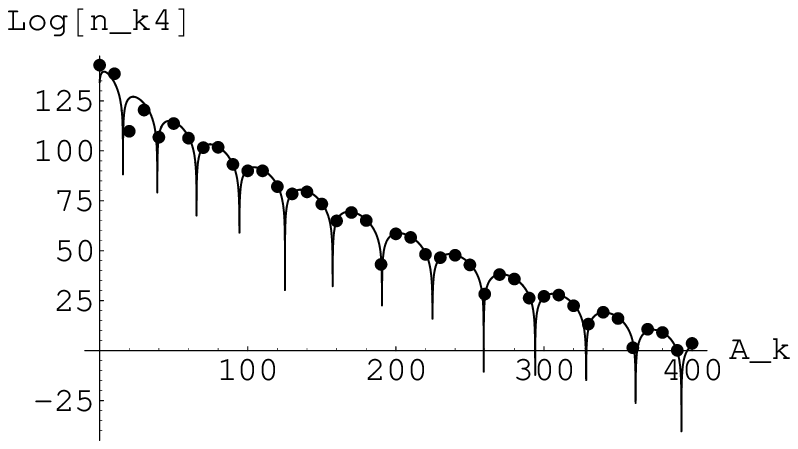}
\end{tabular}
\caption{Occupation number $\log n_4^{(3)}(k)$ as a function of $A_k$, resulting from tachyonic 
resonance due to the interaction $\frac{\phi^3 \chi^2}{M}$ after $j = 4$ inflaton oscillations, for $q_3 = 20$ 
(left) and $q_3 = 200$ (right). The curve is obtained from the analytical formula (\ref{nkjg}) while the dots 
correspond to numerical solutions of (\ref{nmathieu}) with positive frequency initial conditions.}
\label{nk5legs}
\end{center}
\end{figure}

For illustration, we apply the obtained results for a particular non-renormalizable interaction
 $\lambda_3 \frac{\phi^3 \chi^2}{M}$, corresponding to $n=3$ in formulae above.
The spectrum of $\chi$ particles created from tachyonic resonance in this case is shown in Figure~\ref{nk5legs} 
for $q_3 = 20$ and $q_3 = 200$, after $j = 4$ inflaton oscillations. 

Comparing to Figure~\ref{nkq20} for $n = 1$, we see that, for the same not very large value 
of the $q$-parameter ($q = 20$, left panel of 
Figure~\ref{nk5legs}), the WKB approximation is less accurate for $n = 3$ (and more generally for $n > 1$) than for 
$n = 1$. This approximation improves with the growth of $q$.
Recall that for $n=1$ the approximation works perfectly even for moderate $q$.

To understand the subtle difference between the $n=1$ and $n > 1$ cases, we shall
 estimate the range of validity of the WKB approximation for arbitrary $n$. 
Indeed, applying (\ref{valmatch}) to (\ref{nmathieu}), we  derive the range of validity  
for the matching procedure leading to (\ref{nkjg}) (see Appendix A)
\be
\label{valmatchn}
\left[\frac{\left(n-1\right)^3}{16\,q_n}\right]^{n/(n+2)} \ll \; \frac{A_k}{2q_n} < \; 1 - \sqrt{\frac{n}{q_n}} 
\ee
The upper bound was already obtained in 
Section~\ref{tachres} in the particular case $n = 1$. However, for $n \neq 1$, we now have a lower bound too. 
This explains the mismatch between the analytical and numerical results for small $A_k$ in Figure~\ref{nk5legs}  
for $q_3 = 20$ (left panel), which was absent for $n = 1$. The reason why (\ref{nkjg}) is less accurate for 
$A_k \approx 0$ when $n > 1$ is that, in this case, the frequency squared in (\ref{nmathieu}) varies as 
$|z-z_0|^n$ in the vicinity of the turning point $z_0$, whereas the matching procedure used relies on the 
linear approximation (\ref{first}), which breaks down for $A_k \approx
0$ (see Appendix A for details). On the other hand, we see from 
(\ref{valmatchn}) that the range of $A_k$ for which (\ref{nkjg}) is accurate increases with $q$. This is confirmed 
in the right panel of Figure~\ref{nk5legs}. At the beginning of preheating, the $q$-parameter is very large and 
the analytical approximation (\ref{nkjg}) is very accurate. 

We see that the impact of non-renormalizable interactions on preheating can be very strong.

Let us now include the expansion of the universe.
We can repeat the previous calculations just by ``adiabatic'' scaling of parameters
with the scale factor
\begin{equation}
A_k \rightarrow
\frac{A_k}{a^2} \ , \,\,\,\,q_n \rightarrow \frac{q_n}{a^{3n/2}}
\label{nonrenends}
\end{equation}

For non-renormalizable interactions, the $q$-parameter decreases with time faster for bigger $n$, 
and in particular faster than for the trilinear interaction. For $n > 1$, it decreases faster 
than the square of the physical momenta, which means that in this case fewer and fewer modes are amplified 
by the tachyonic effect. The overall effect of the 
expansion of the universe is thus to decrease the effectiveness of
tachyonic resonance (see (\ref{Xknap})), and this occurs faster for non-renormalizable 
interactions than for the trilinear one. In particular, depending on the interactions 
which stabilize the potential, the preheating stage may now terminate in the regime of 
narrow parametric resonance, $q_n \sim 1$ (see Section (\ref{tachres})). 
For example, for $n=9$ and $q_n \sim 10^{12}/10^n$ at the beginning of preheating, this happens 
after only $2$ oscillations of the inflaton field.

Let us now consider other terms in the potential 
of the effective field theory describing the 
interactions between the inflaton and the matter field $\chi$, such as a four-legs interaction and 
non-renormalizable terms $\lambda_n \frac{\phi^n \chi^2}{M^{n-2}}$ with different $n$. The effective 
mass squared for $\chi$ is given by 
$$
m_{\chi, \mathrm{eff}}^2 = m_{\chi}^2 + \sigma\,\phi + g^2\,\phi^2 + \frac{\lambda_3}{M}\,\phi^3 + \frac{\lambda_4}{M^2}\,\phi^4 + ...
$$
As we already discussed, the contribution of the renormalizable interactions is generally dominated by the term  
$g^2\phi^2\chi^2$ in the potential. At the beginning of preheating, the amplitude of the inflaton is given by 
$\Phi \sim 0.1\,M_P$. For $g^2 < \frac{\Phi}{M} < 1$ and all the couplings $\lambda_i$ of order one, we see 
that preheating is dominated by the term $\lambda_3 \frac{\phi^3 \chi^2}{M}$ in the potential. The 
important conclusion is that it dominates over the $\lambda_4 \frac{\phi^4 \chi^2}{M}$ term, the 
$\lambda_5 \frac{\phi^5 \chi^2}{M}$ term, and the other
non-renormalizable terms, as well as over the renormalizable terms such as $g^2 \phi^2 \chi^2$.

\section{Conclusions}\label{conclusions}

The main purpose of this paper is the study of the impact of bosonic trilinear interactions
on the decay of the inflaton after chaotic inflation, and on the following thermalization stage.
The starting motivation is that terms of the form $\phi \chi^n$ (where $\phi$ is the
inflaton, while $\chi$ represents some matter field) are necessary for a complete
decay of the inflaton, and the trilinear interaction $\phi \chi^2$ is certainly the most
common of such interactions. In particular, it appears naturally in supersymmetric theories. 
In our analysis, we found that a trilinear coupling can lead to other relevant effects.

One of them is the appearance of a qualitatively new type of preheating, for which we developed 
an analytical theory. On the formal side, we extended  the 
theory of broad parametric resonance, developed in~\cite{KLS97} for the region $A > 2q$, to the 
remaining portion of parameters where $0 < A < 2q$. Preheating in this
region can be best understood as a combination of parametric resonance
and tachyonic effects, so we called it ``tachyonic resonance.'' For
half of its oscillation the inflaton is negative and $\phi \chi^2$ provides a tachyonic effective 
mass for the scalar $\chi \,$. This tachyonic instability recurs
periodically in time, which results in the presence of resonance/stability bands, which is 
typical of parametric resonance. Depending on the relative strengths
of the interactions, this
effect can be dominant or subdominant relative to the more studied parametric resonance from a 
quartic $\phi^2 \chi^2$ interaction. If the trilinear interaction comes from a simple superpotential, the quartic
interaction is dominant at early times. Nevertheless, the trilinear interaction tends to enhance the
parametric resonance driven by the four-legs interaction,  due to the fact that the resonance
occurs when $\phi \approx 0 \,$, where the relative contribution of the cubic over the quartic interaction 
is maximal. Even if initially subdominant, the cubic interaction eventually comes to dominate over the quartic 
one due to the decrease of $\phi$ as the universe expands. Eventually, it leads to a complete decay of the
inflaton, while the quartic interaction freezes out.

A different noticeable effect of the trilinear term is a speed up of the thermalization of the system.
The distributions produced at preheating are largely populated at (relatively) low momenta, and they are much 
more enhanced in the infrared with respect to a
thermal distribution. Therefore, thermalization proceeds through particle fusion. As can be expected, 
the presence of a trilinear interaction enhances this effect. We
confirmed this by evolving fields on the lattice and comparing models with and without the $\phi \chi^2$ term. 
If only the quartic $\phi^2 \chi^2$ interaction is present,
preheating is followed by a turbulent rescattering stage, where quanta of $\phi$ are also excited, and 
by a much longer phase where the distributions very slowly
evolve towards the ultraviolet. In the meantime, the equation of state (EOS) of the system rises to an 
intermediate value between the one of matter and radiation, and
it then evolves back towards the one of matter, due to the increasing importance of nonrelativistic quanta of 
the massive inflaton. On the contrary, in the cases in which the trilinear term was present we noticed a much 
quicker population of the UV modes, particularly for the inflaton field. As a consequence, the EOS remains fixed at the
intermediate level throughout our simulation, while the fraction of relativistic quanta of $\phi$ continues to 
increase. Although thermalization completes on a much longer timescale
than we can simulate, details of this early stage can be extremely 
relevant for the production of heavy particles, gravitational relics, and for modulated
perturbations, as we have described in~\cite{podol}.

Tachyonic resonance (with small variations of details) also arises 
for non-renormalizable interactions in $\phi^n\,\chi^2$ with $n>1$ odd, and we
extended our analytical formalism to such couplings.  
Non-renormalizable interactions naturally arise in 
supergravity models, where they are suppressed by powers of the Planck mass. In chaotic inflation, $\phi \sim 0.1
M_p \,$ at the beginning of reheating, so the higher order operators are progressively more and more suppressed 
(this guarantees that the computation is under control). However, quite surprisingly, we found that a  
$\lambda \, \phi^3\,\chi^2 / M_p$ term (with $\lambda$ of order one) dominates over the quartic $g^2 \phi^2\,\chi^2$ 
interaction at the beginning of preheating, for the values of coupling $g^2$ which are typically 
considered. Such a $\lambda \, \phi^3\,\chi^2 / M_p$ term leads also to much stronger preheating, 
via tachyonic resonance. This opens a qualitatively new possibility for inflaton
decay, even if it interacts only very weakly with 
the Standard Model sector through usual renormalizable
interactions.

\section{Acknowledgements}

It is a pleasure to thank Andrei Linde, Wan-Il Park, and Ewan Stewart for useful discussions. 
G.F. and M.P. would like to thank CITA for
their hospitality during this work. L.K. was supported by NSERC and CIAR.
G.F. was supported by
NSF grant PHY-0456631. M.P. was supported in part by the DOE grant
DE-FG02-94ER40823, and by a grant from the Office of the Dean at the Graduate School of the University of Minnesota.

\appendix
\section{Derivation of the Transfer Matrix (\ref{transmat}) for Tachyonic Resonance}

In this appendix, we outline the procedure to derive the transfer matrix (\ref{transmat}) and 
we discuss its range of validity.\\

First we want to follow the evolution of the modes from an oscillatory
form (\ref{chiom}) to an exponential form (\ref{chiOm}) 
in the vicinity of the turning point $t = t_{kj}^-$. Let us decompose the frequency to linear order
\be
\label{first}
\omega_k^2 = -\Omega_k^2 \simeq \frac{d\omega_k^2}{dz}\left(t_{kj}^-\right)\;(t_{kj}^- - t) \; .
\ee
Along the real axis of  $t$ we can only use the WKB approximation far away from the turning point.
However, we can implement here the following trick known in quantum mechanics \cite{LL3}.
 By analytical continuation of $\chi_k(t)$ to the complex plane, we may pass around the turning point along 
a path in the complex plane $t$ which is sufficiently far from it so that the WKB approximation remains valid, but also 
sufficiently close to it so that we may use (\ref{first}). It follows that 
\be
\label{valmatch}
\left[4\,\frac{d\omega_k^2}{dz}\left(t_{kj}^-\right)\right]^{-1/3} \ll \; |t-t_{kj}^-| \; \ll  
2\;\frac{d\omega_k^2}{dz}\left(t_{kj}^-\right) \; \left[\frac{d^2\omega_k^2}{dz^2}\left(t_{kj}^-\right)\right]^{-1}
\ee
has to be satisfied along the contour. For the three-legs interaction (\ref{mathieu}), such a contour always exists 
for the modes with $A_k < 2q - 2\sqrt{q}$, which is in any case required for the approximation 
(\ref{chiOm}) to be valid (see section 3). The constraint (\ref{valmatch}) is actually more severe in the case of 
non-renormalizable interactions, as discussed in Section \ref{nonrenorm}.

\begin{figure}[hbt]
\includegraphics[width = 10 cm]{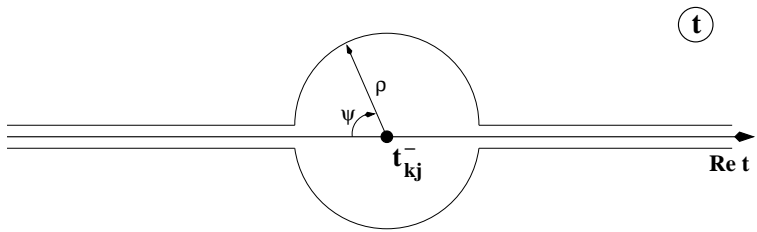}
\caption{For sufficiently large $\rho$, the WKB approximation is applicable everywhere along both contours.}
\label{contours}
\end{figure}

The contour may be chosen  in the lower or 
in  the upper half of the complex plane as shown in  Figure~\ref{contours}.
Using the representation $t_{kj}^- - t = \rho\,e^{i\psi}$ and varying $\psi$ from $0$ to $\pi$, 
one goes from negative to positive $t - t_{kj}^-$ along a contour in the lower half of the complex 
plane. This gives
\be
(t_{kj}^- - t)^{1/2} \, \rightarrow \, (t - t_{kj}^-)^{1/2}\,e^{i\pi/2}
\ee
so that (\ref{chiom}) is matched to 
\be
\label{lower}
\chi_k^j \, \rightarrow \, \frac{\alpha_k^j\,e^{-i(\theta_k^j + \pi/4)}}{\sqrt{2\Omega_k}}\;
e^{\int_{t_{kj}^-}^t dt\,\Omega_k(t)}
\ee
where we neglected an exponentially small term proportional to $\beta_k^j$, which is beyond the 
accuracy of the WKB approximation. With our convention, $\theta_k^j$ is the total phase accumulated 
from $t_0$ to $t_{kj}^-$ during the intervals where $\omega_k^2 > 0$, 
$\theta_k^j = \int_{t_0}^{t_{kj}^-} dt\,\omega_k(t)$. Similarly, going along a contour in 
the upper half of the complex plane gives
\be
\label{upper}
\chi_k^j \, \rightarrow \, \frac{\beta_k^j\,e^{i(\theta_k^j + \pi/4)}}{\sqrt{2\Omega_k}}\;
e^{\int_{t_{kj}^-}^t dt\,\Omega_k(t)}
\ee
Comparing (\ref{lower}) and (\ref{upper}) to (\ref{chiOm}), one finds
\be
\label{tk-}
b_k^j = \alpha_k^j\,e^{-i(\theta_k^j + \pi/4)} + \beta_k^j\,e^{i(\theta_k^j + \pi/4)}  
\ee
which relates the evolution of the modes before and after the turning point $t = t_{kj}^-$. 
Repeating the same procedure around the turning point $t = t_{kj}^+$ gives 
\be
\label{tk+}
\beta_k^{j+1} = e^{X_k^j}\,e^{-i(\theta_k^j + \pi/4)}\,b_k^j \;\;\; \mbox{ and } \;\;\; 
\alpha_k^{j+1} = e^{X_k^j}\,e^{i(\theta_k^j + \pi/4)}\,b_k^j
\ee
where $X_k^j$ is defined in (\ref{Xkj}). Again, the matching procedure is accurate if the contour may be 
chosen such that a relation similar to (\ref{valmatch}) is satisfied along it. 
The transfer matrix (\ref{transmat}) follows directly 
form (\ref{tk-}), (\ref{tk+}).

A similar result can be obtained if we substitute the linear form (\ref{first}) in the wave equation (\ref{modes})
and observe that its solutions are given by Airy functions.
Matching asymptotics of the Airy functions with the oscillating  (\ref{chiom}) and exponential forms  (\ref{chiOm})
gives the transfer matrix (\ref{transmat}). 

\section{Comments on Stability Bands}

In the regime of tachyonic resonance in the region where $A < 2q$ the stability bands are very narrow  for $q \gg 1$.
 The situation is reversed with respect to conventional parametric
 resonance for $A > 2q$ where for $q \ll 1$ the
instability bands are very narrow.
Formally, there is an interesting analogy between time-dependent solutions of the wave equations with periodic frequency
and the one-dimensional Schr\"{o}dinger equation for $x$-dependent
 wave function in a spatially periodic potential.
Solving the eigenvalue problem for the Schr\"{o}dinger equation with a periodic potential results in finding 
the bound states and their  energy levels. Each energy level is slightly washed out due to the tunneling through the barriers
between minima of the potential. Those energy bands exactly corresponds to the stability bands in the theory
of time-dependent tachyonic resonance.

Let us first  consider the Schr\"{o}dinger equation  with the potential $V(z) = 2q \cos 2z$ for $0<z<\pi$,
so that it has only one minimum.
In the WKB approximation the bound energy  levels are defined with  the Bohr-Zommerfeld quantization formula 
$\Theta_k (A_k, q)=\pi (n+1/2)$,
where $\Theta_k$ is the phase of the wave function and $n$ is an integer number.  In our case we have 
\be
\frac{\Theta_k (A_k, q)}{\pi} = \frac{1}{\pi} \int_{\hlf {\rm arccos} \frac{A_k}{2q}}^{\pi - \hlf {\rm arccos} \frac{A_k}{2q}}
\sqrt{A_k - 2q \cos 2z}\,dz = n + \hlf
\label{BZ1}
\ee
In the time-dependent problem of tachyonic resonance this exactly corresponds to the stability band around $k$ where 
$\cos \Theta_k (z) = 0$, see (\ref{nk0}). 

The function $\Theta_k (A_k, q)$ was calculated in (\ref{tthetak}). 
To find the position of the stability bands in the $(A_k, q)$ plane, we have to invert 
(\ref{BZ1}). We  notice, by direct differentiation,  that 
\be
\frac{\partial \Theta_k (A_k, q)}{\partial A_k} = \frac{T}{4}
\label{diff}
\ee
where $T$ is the classical period of motion between the turning points in the potential,
\be 
T = 2\int_{\hlf {\rm arccos} \frac{A_k}{2q}}^{\pi - \hlf {\rm arccos} \frac{A_k}{2q}} \frac{dz}{\sqrt{A_k - 2q \cos 2z}} \ .
\label{period1}
\ee
Using properties of elliptic integrals, this reduces to
\be
T = \frac{2}{\sqrt{q}}\;K\left(\frac{A_k + 2q}{4q}\right)\;,
\label{period}
\ee
where $K$ is the complete elliptic integral of the first kind, which has a logarithmic divergence when its argument 
tends to $1$ (see~\cite{abram}). We find the accurate approximation in the range $0 \leq A_k < 2q$
\be
\label{approxT}
T \simeq \frac{3.71}{\sqrt{q}}\,\left[1-0.235\,\ln\left(1-\frac{A_k}{2q}\right)\right] 
\ee

Integrating (\ref{diff}) then gives
\be
\label{approxttheta}
\Theta_k (A_k, q) = \sqrt{q}\,\left[a + b\,\frac{A_k}{2q} + c\,\left(1-\frac{A_k}{2q}\right)\,\ln\left(1-\frac{A_k}{2q}\right)\right]
\ee
A very good approximation is obtained for $a=1.69$, $b=2.31$ and $c=0.46$. Using the asymptotic behavior of 
(\ref{approxttheta}) for $A_k/2q \rightarrow 0$ and $A_k/2q \rightarrow 1$, we find from (\ref{BZ1}) the distance 
between the adjacent stability bands:
\be
\label{interstab}
\Delta A_k = d\,\sqrt{q}
\ee
where $d\simeq 3.41$ for $A_k/2q \rightarrow 0$ and $d\simeq 2.73$ for $A_k/2q \rightarrow 1$. 
The stability bands are approximately equidistant in both limits. Eq.~(\ref{interstab}) agrees  
with Fig.~\ref{nkq20} (right panel). 

Using the analogy with quantum mechanics, we can go \textit{beyond} the WKB approximation. In particular, one can 
take the periodicity of the quantum mechanical potential into
account~\cite{LL9} and find the exact wave function of the particle $\chi_k$, which will
have the form of a Bloch wave function. It turns out that the only effect of the periodicity of the potential on the spectrum 
is the broadening of levels (\ref{BZ1}). The characteristic width is 
\be
\delta A_k \sim \frac{\sqrt{D}}{T},
\label{broadening}
\ee
where $D = \exp \lmk - X_k \rmk$ is the transmission coefficient and
the function $X_k$ was calculated in (\ref{Xk}, \ref{appXk}). It follows that, to first order in $A_k/2q$, 
the width of the bands varies with $q$ as $\delta A_k \sim \sqrt{q}\,e^{-x\sqrt{q}}$, where $x \simeq 0.85$. 

The stability/instability chart of the Mathieu equation for a time-dependent wave function
tells us which bands are stable and which are unstable,
but it does not address the question of how these bands are populated with time.
The analogy with the quantum mechanical problem of a spatially
periodic potential gives us insight into how 
it occurs. After the first background oscillation the $\chi$ waves  are not amplified at the
discrete modes $k$ which experience destructive interference, corresponding to the
quantization  (\ref{BZ1}). After the second oscillation the non-amplified modes split, and this splitting further continues
as the number of oscillations accumulate. Asymptotically the splitting (non-amplified) levels
cover a gap which corresponds to the stability band.


\end{document}